%% file: IJOME_Mycek_et_al_2013.tex
\documentclass[preprint, onecolumn, 10pt]{elsarticle}
\journal{International journal of Marine Energy}

\usepackage[utf8]{inputenc}
\usepackage[T1]{fontenc}
\usepackage[english]{babel}

\usepackage{geometry}

\biboptions{square, comma, numbers, sort&compress}

\usepackage{graphicx}
\usepackage{color}

\usepackage{amsmath}
\usepackage{amsfonts}
\usepackage{amssymb}

\usepackage{caption}
\usepackage{subfigure}

\usepackage[colorlinks=true]{hyperref}
\usepackage[all]{hypcap}
\usepackage{url}

\usepackage{array}

\input{math-preamble.tex}

\newcommand{\up}[1]{\textsuperscript{#1}}

\begin{document}

\renewcommand{\today}{~}

\input{frontmatter.tex}

\input{abstract.tex}
\input{intro.tex}

\input{single.tex}
\input{array.tex}

\input{concl.tex}
\input{acknow.tex}

\bibliographystyle{model1-num-names}

\end{document}

%% file: math-preamble.tex
\usepackage{amsmath}
\usepackage{amssymb}
\usepackage{amsfonts}
\usepackage{amsthm}

\usepackage{textcomp}

\usepackage{empheq}

\renewcommand\d{\mathrm d}

\renewcommand{\vec}[1]{\ensuremath{\mathbf{#1}}}

\newcommand{\vU}{\vec{U}}

\newcommand{\mc}[1]{\mathcal{#1}}

%% file: frontmatter.tex
\begin{frontmatter}



\title{Numerical and Experimental Study of the Interaction between two Marine Current Turbines}


\author[lomc,ifremer]{Paul Mycek}
\ead{paul.mycek@ifremer.fr}

\author[ifremer]{Beno\^it Gaurier}
\ead{benoit.gaurier@ifremer.fr}

\author[ifremer]{Gr\'egory Germain}
\ead{gregory.germain@ifremer.fr}

\author[lomc]{Gr\'egory Pinon}
\ead{gregory.pinon@univ-lehavre.fr}

\author[insa,lomc]{Elie Rivoalen}
\ead{elie.rivoalen@insa-rouen.fr}

\address[lomc]{Laboratoire  Ondes   et  Milieux  Complexes,   UMR  6294,
  CNRS--Universit\'e du Havre,\\%
  53, rue de Prony, BP 540,\\%
  F-76058 Le Havre Cedex, France.}
\address[ifremer]{IFREMER, Marine Structures Laboratory,\\%
  150, quai Gambetta, BP 699,\\%
  F-62321 Boulogne-Sur-Mer, France.}
\address[insa]{Laboratoire d'Optimisation  et Fiabilit\'e en M\'ecanique
  des Structures, EA 3828, INSA de Rouen,\\%
  Avenue de l'Universit\'e, BP 08,\\%
  F-76801 Saint-Etienne-du-Rouvray, France.}

\input{abstract.tex}

\end{frontmatter}

%% file: abstract.tex
\begin{abstract}
  The  understanding  of   interaction  effects  between  marine  energy
  converters  represents the  next  step in  the  research process  that
  should eventually  lead to the  deployment of such  devices.  Although
  some  \textit{a priori} considerations  have been  suggested recently,
  very few real condition studies  have been carried out concerning this
  issue.

  Trials were run on 1/30\up{th} scale models of three-bladed marine
  current turbine prototypes in a flume tank.  The present work focuses
  on the case where a turbine is placed at different locations in the
  wake of a first one.  The interaction effects in terms of performance
  and wake of the second turbine are examined and compared to the
  results obtained on the case of one single turbine. Besides, a
  three-dimensional software, based on a vortex method is currently
  being developed, and will be used in the near future to model more
  complex layouts.

  The  experimental  study  shows  that  the second  turbine  is  deeply
  affected by the  presence of an upstream device  and that a compromise
  between  individual  device performance  and  inter-device spacing  is
  necessary.  Numerical results show  good agreement with the experiment
  and are promising for the future modelling of turbine farms.
\end{abstract}


%% file: intro.tex
\section{Introduction}
A new level has been recently reached in the deployment of marine energy
converters with the launching of several large-scale projects.  For
instance, India plans to install a 250MW tidal farm on its west coast,
with a first implantation of fifty AK1000 (\textit{Atlantis Resources})
turbines of 1MW each~\cite{BBC_India2011}. In 2011, the Scottish
Government announced its approval for a 10MW tidal power array project
on Scotland's west coast.  It would consist of ten HS1000
(\textit{Hammerfest Strøm}) turbines that should be installed from 2013
to 2015~\cite{ScottishPower_Renewables2011}.  \textit{Marine Current
  Turbines Ltd} also announced in March 2011 that it had submitted a
consent application to install in 2015 a 10MW tidal farm off the
Anglesey coast, in Wales~\cite{MCT2011}.  Some \textit{a priori} studies
have already been carried out to evaluate the potential retrievable
power in specific areas, for instance the Race of Alderney, off the Cap
de la Hague in France~\cite{Bahaj2004}.

The behaviour of single marine  energy converters such as marine current
turbines  is now  globally well  understood thanks  to  experimental and
numerical studies~\cite{Batten2008, Baltazar2008, Maganga2010}. However,
as the size of  such arrays is expected to grow with  time, the issue of
interaction  effects between turbines,  most importantly  negative ones,
has  to  be  addressed.    Some  \textit{a  priori}  considerations  and
suggestions  have  been  presented  in  recent  studies~\cite{Myers2010,
  Rawlinson-Smith2010}  about different parameters  of a  marine current
turbines  array  layout.   As  regards  interaction  effects,  numerical
studies  on vertical  axis tidal  turbines have  been carried  out, such
as~\cite{Li2011} about the torque fluctuation.

The  aim of the  present paper  is to  give an  idea of  the interaction
characteristics  between two  horizontal axis  current turbines  in real
condition  configurations.   It   complements  a  previous  experimental
study~\cite{Maganga2010}   carried  on   different  but   similar  blade
geometries and additional rotation speeds.  Another purpose is to validate the
three-dimensional  software   on  a  single   device  configuration,  before
extending it to multi-devices cases in  the future. The main point is to
be able  to model more complex  turbines array layouts,  which cannot be
set up in flume tanks.

The first part of this paper is thus dedicated to the characterisation
of a single marine current turbine behaviour and to the validation of
the numerical software on these cases.  The second part presents
experimental trials on two-device configurations and uses the results on
single device configurations as a comparison. Conclusions can then be
drawn on the interaction effects between two marine current turbines. A
comparison with the latest numerical results is also briefly presented.
Eventually, conclusions are drawn and an outlook on both numerical and
experimental future work is given.


%% file: single.tex
\section{Single-device configuration}
\label{sec:single}

Trials  have  been  performed  at  the  French  Research  Institute  for
Exploitation  of the Sea  (IFREMER) on  a 1/30\up{th}  scale model  of a
three-bladed turbine prototype,  in a 18$\times{}$4$\times{}$2m wave and
current flume tank.  The prototype  consists of a rotor, which is $0.7$m
in    diameter,     and    a     $0.7$m    long    axial     hub    (cf.
figure~\ref{fig:single_param}).   The blockage ratio  is then  less than
5\%. The turbine blades are designed from a NACA63418 profile.

Force and moment on the turbine are measured thanks to six-component
load cells, while velocity measurements are obtained thanks to a two
component Laser Doppler Velocimetry (LDV) system.  The experimental
setup is presented in details in~\cite{Maganga2010}. In this section,
experimental trials, concerning the wake behind the turbine and the
performance of the device, are presented. Corresponding numerical
results are shown in order to validate the numerical tool.

\subsection{Description of the parameters}
\label{sec:single_param}
\begin{figure}[ht]
  \centering
  \resizebox{0.7\linewidth}{!}{\large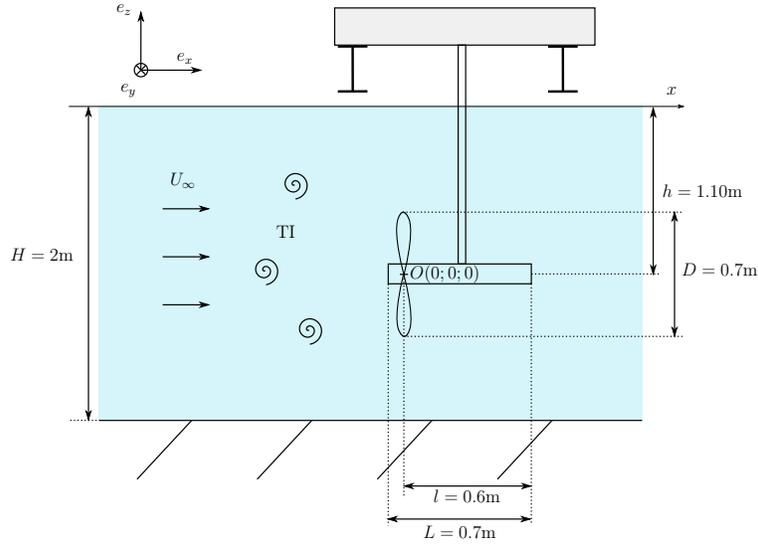}
  \caption{Description  of  the  single  configuration  parameters.  The
    origin $O(0;0;0)$ is chosen at the rotor centre.}
  \label{fig:single_param}
\end{figure}

A marine current  turbine may be subject to  various parameters that can
influence its behaviour, amongst others:
\begin{itemize}
\item  The  current velocity  $\vU_{\infty}$  which  is assumed to  be
  uniform and such that
  \begin{equation}
    \vU_{\infty}=U_{\infty}\vec{e}_x
  \end{equation}
\item The Tip Speed Ratio (TSR):
  \begin{equation}
    \label{eq:TSR}
    TSR=\frac{\Phi R}{U_{\infty}}
  \end{equation}
  where $R=D/2$ denotes  the rotor radius and $\Phi$  the rotation speed
  of the turbine.
\item The ambient turbulence intensity rate (TI) defined by
  \begin{equation}
    \label{eq:TI}
    TI=100\frac{\sqrt{\frac{1}{3}(\sigma_u^2+ \sigma_v^2 + \sigma_w^2)}}
    {\sqrt{\bar{u}^2+\bar{v}^2+\bar{w}^2}}
  \end{equation}
  where $u$, $v$ and $w$ are  respectively the $x$, $y$ and $z$ velocity
  components,  $\bar{q}$ denotes  the  mean value  and $\sigma_{q}$  the
  standard deviation of quantity $q$.
\end{itemize}
These parameters, as  well as the flume tank  and turbine geometries are
summarized on figure~\ref{fig:single_param}.

In the present study, an incoming velocity of 0.8m/s is considered, as well as TSR values between
0 and 10, and ambient TI of 3\% and 15\%.  The turbulence in the flow is
induced by the current generator of the flume tank.  Without the use of
a honeycomb, a turbulence intensity rate of 15\% is measured, which can
be reduced to 3\% by placing honeycomb grids at the beginning of the
flume.  LDV Measurements, performed at different locations in the area
swept by the turbine blades, indicate that these sources of turbulence
imply an homogeneous turbulence structure.

\subsection{Characterization of the wake}
\label{sec:single_wake}


The general  aspect of the wake  behind a marine current  turbine can be
estimated  by drawing  velocity and  turbulence maps  of  the downstream
flow.  Such maps  can  provide useful  information  to characterize  the
impact of the turbine on its close environment.

The LDV measurements are performed on a grid whose nodes $(X_i,Y_i)$ are
arranged as follows:
\begin{itemize}
  \item $X_1=1.2D$ and $X_i=i \times D$ for $i=2,\ldots,10$.
  \item  $Y_i=-1.2 + (i-1)  \times 0.1$m  for $i=1,\ldots,25$,  with two
    additional positions $Y_{26}=-Y_{27}=R=0.35$m;
\end{itemize}
The measurement  on each  node lasts 100  seconds with an  observed data
rate between 7 and 17Hz.

The  downstream  turbulence $TI_{down}$  is  evaluated  as a  turbulence
intensity rate in the $xOy$ plan:
\begin{equation}
  \label{eq:TI_down}
  TI_{down}=100\frac{\sqrt{\frac{1}{2}(\sigma_u^2+ \sigma_v^2)}}
  {\sqrt{\bar{u}^2+\bar{v}^2}}
\end{equation}
As for  the axial velocity,  $u/U_\infty$ is considered,  which represents
the proportion of velocity recovered behind the turbine in comparison to
the upstream velocity $U_\infty$.

\begin{figure}[!ht]
  \centering
  \subfigure[Axial velocity map (TI=3\%)\label{fig:wake_1hydrol_TSR3_67_vit5}]{\includegraphics[width=0.48\linewidth]{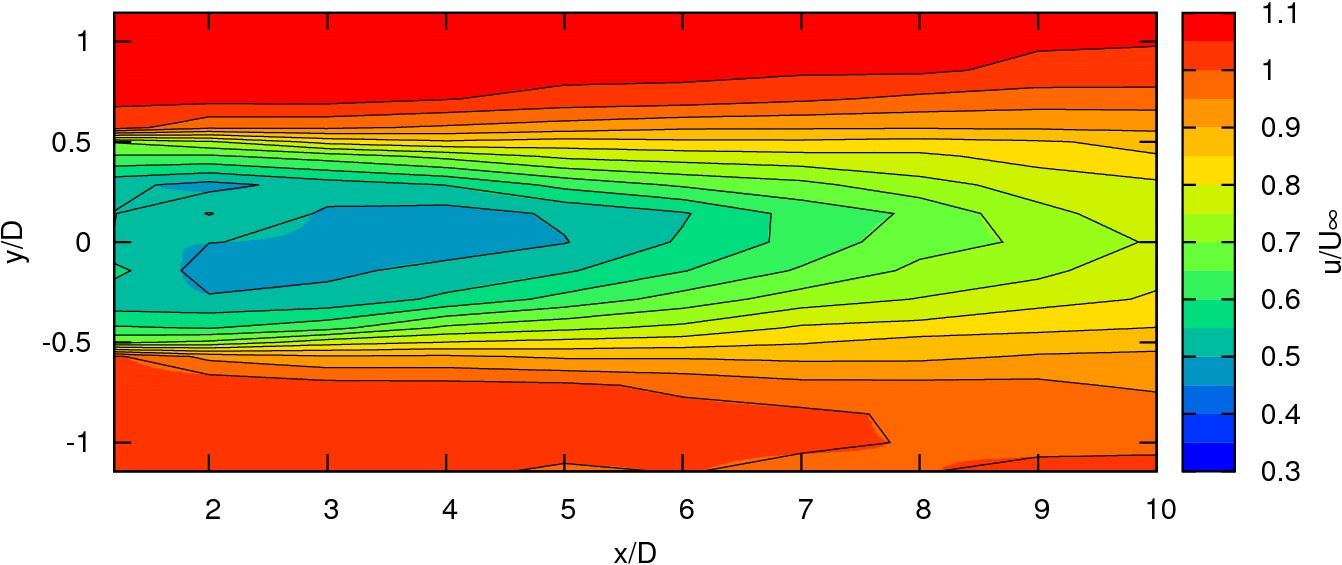}}
  \subfigure[Axial velocity map (TI=15\%)\label{fig:wake_1hydrol_TSR3_67_vit25}]{\includegraphics[width=0.48\linewidth]{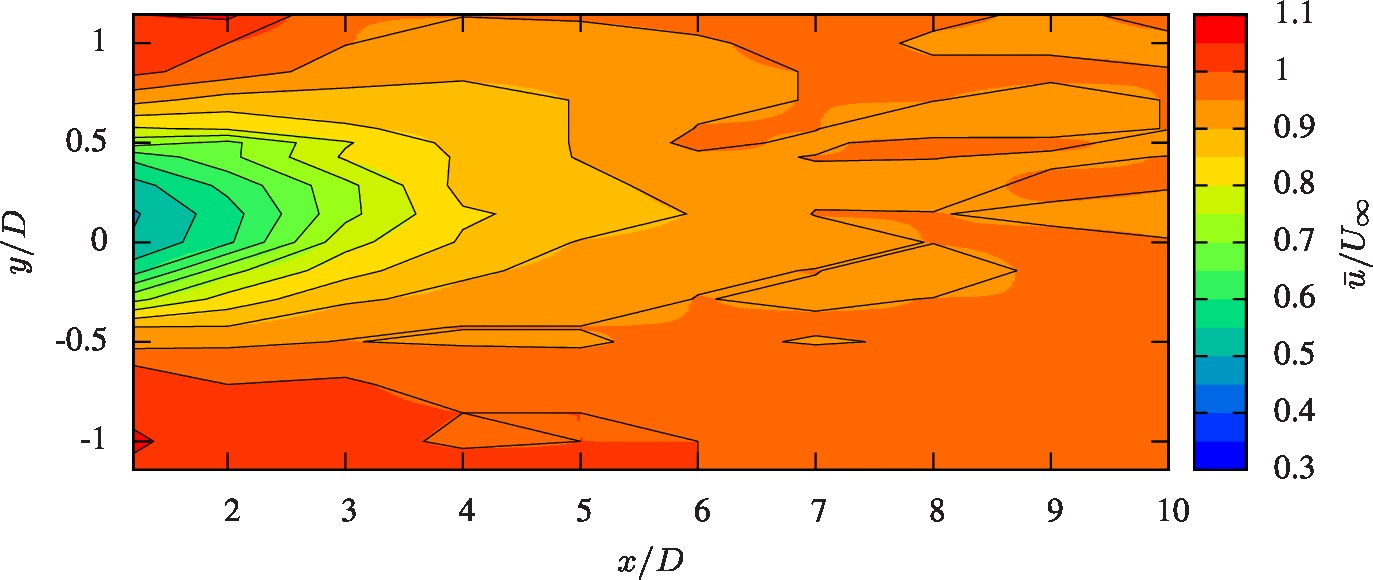}}
  \subfigure[Turbulence map (TI=3\%)\label{fig:wake_1hydrol_TSR3_67_turb5}]{\includegraphics[width=0.48\linewidth]{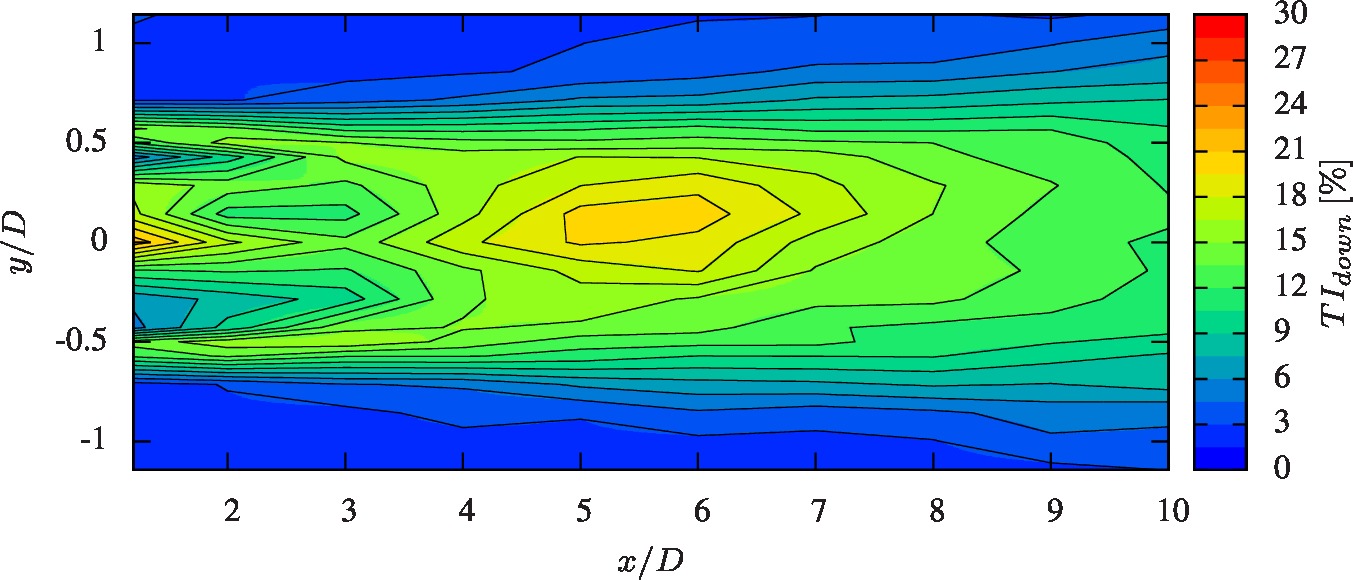}}
  \subfigure[Turbulence map (TI=15\%)\label{fig:wake_1hydrol_TSR3_67_turb25}]{\includegraphics[width=0.48\linewidth]{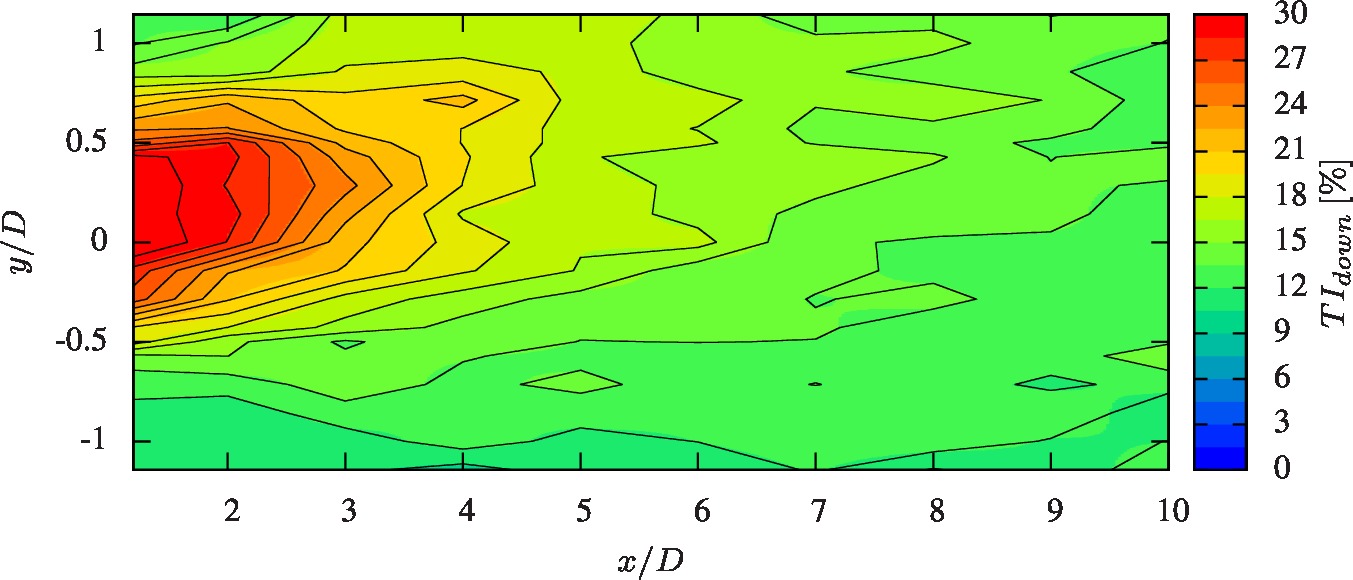}}
  \caption{Wake behind a turbine with TSR=3.67}
  \label{fig:wake_1hydrol_TSR3_67}
\end{figure}

Figure~\ref{fig:wake_1hydrol_TSR3_67} shows axial velocity and
turbulence maps for different upstream ambient TI with a turbine at
TSR=3.67. These maps point out that a higher ambient TI rate reduces the
wake length in terms of velocity and turbulence.  Indeed,
figure~\ref{fig:wake_1hydrol_TSR3_67_vit25} shows that at a distance of
seven diameters behind the turbine in a 15\% ambient TI, the axial
velocity profile tends to recover its uniformity and about 90\% of its
intensity. On the other hand, the profile at the same location for an
ambient 3\% TI (figure~\ref{fig:wake_1hydrol_TSR3_67_vit5}) remains very
wake-shape like, with only 65\% of $U_\infty$ recovered at the centre.
Even 10 diameters behind the turbine, the profile is still non-uniform
and below the oncoming velocity, with about 75\% recovered at the
centre.  The same behaviour can be observed with the downstream
turbulence.  Indeed, with a 3\% ambient TI,
figure~\ref{fig:wake_1hydrol_TSR3_67_turb5} shows that $10$ diameters
behind the turbine, the downstream TI remains higher than 3\%. On the
contrary, it goes back to 15\% in the case of an upstream 15\% TI
(figure~\ref{fig:wake_1hydrol_TSR3_67_turb25}).

These experiments have been performed in the same conditions as
in~\cite{Maganga2010} but with a lower TSR to be able to compared them
with  numerical computations~\cite{Pinon2012}.  As a matter of fact,
as stated in section~\ref{sec:single_comp_num}, computations with too
high TSR are not valid because of the particles emission model.  Another
significant difference resides in the use of  new blades that are not
patented, with a shorter chord, so as to avoid
confidentiality restrictions.

\subsection{Comparison with numerical results}
\label{sec:single_comp_num}

A three-dimensional software is under development in the LOMC laboratory
of Le Havre University, based on a Lagrangian vortex particle
method~\cite{Leonard1980, Winckelmans1993, Cottet2000, Zervos1988,
  Huberson2008}. The vortical flow is discretized into particles, which
are small volumes of fluid carrying intrinsic physical quantities such
as their position and vorticity.  Those particles are emitted at the
trailing edge of the obstacle according to the Kutta-Joukowski condition
and then advected in a Lagrangian frame thanks to the Navier-Stokes
equations for an incompressible flow. The details of the method
are presented in~\cite{Pinon2005a, Pinon2012}.

\begin{figure}[ht]
  \centering
  \includegraphics[width=0.48\linewidth]{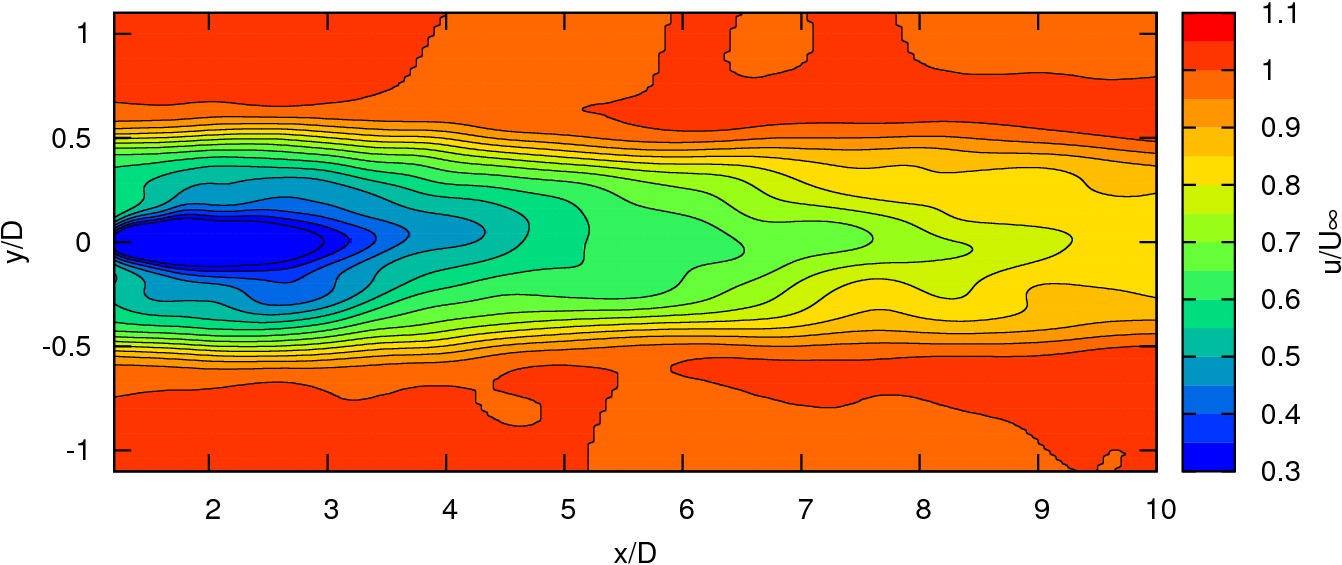}
  \caption{Axial   velocity   map   from   computation   with
    TSR=3.67, TI=0\% and a turbulence model.}
  \label{fig:wake_1_hydrol_TSR3_67_num}
\end{figure}

Velocity  maps  can  also  be  drawn from  the  numerical  computations.
Figure~\ref{fig:wake_1_hydrol_TSR3_67_num}  presents  an axial  velocity
map obtained  from a  numerical computation and  shows that  the general
aspect of the wake is well reproduced. It should be pointed out that all
of the  computations are performed with  a 0\% ambient TI,  with a Large
Eddy Simulation (LES) turbulence model.


\begin{figure}[ht]
  \centering
  \includegraphics[width=0.6\linewidth]{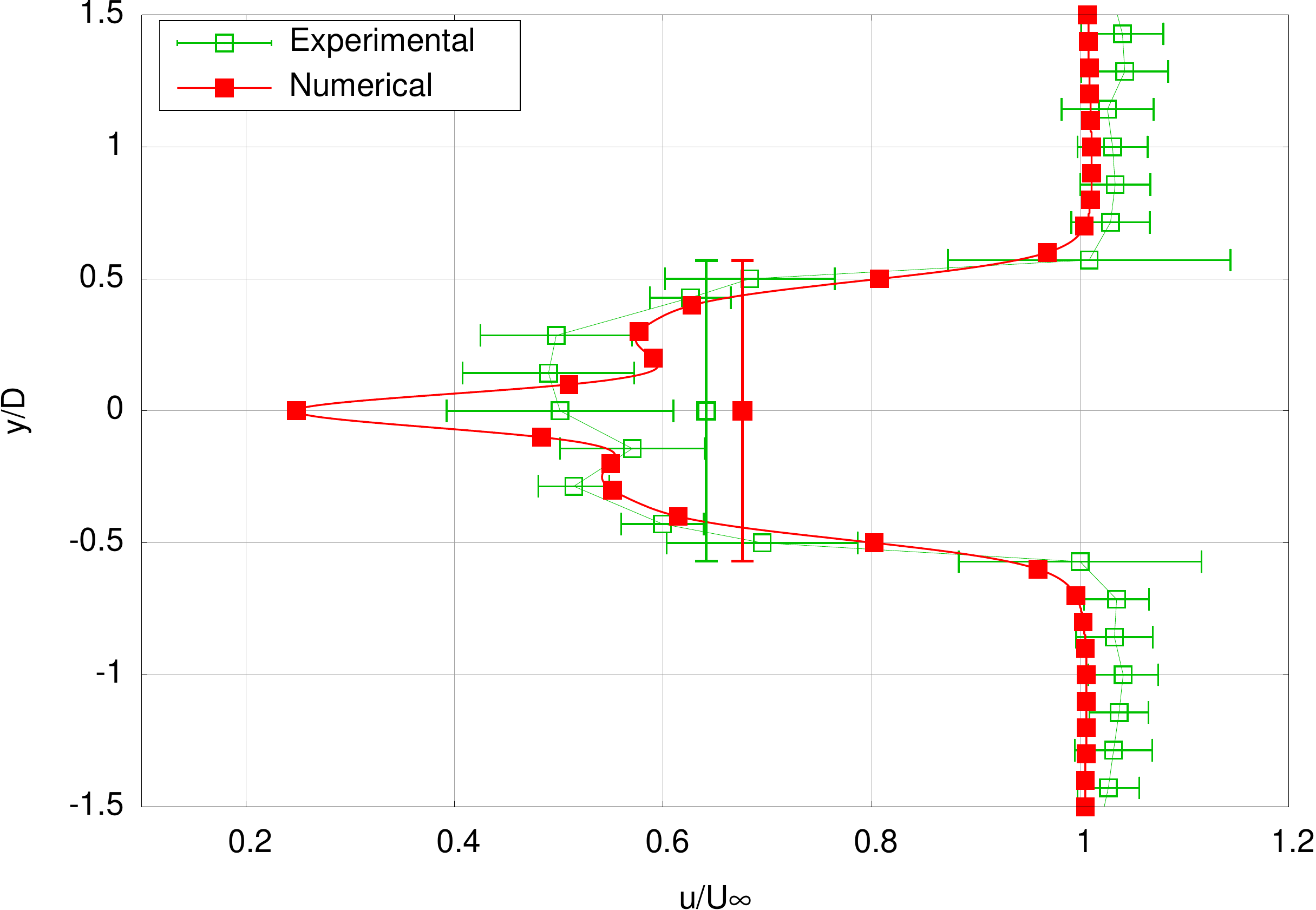}
  \caption{Numerical  (TI=0\% with  turbulence  model) and  experimental
    (TI=3\%) axial  velocity profile with TSR=3.67.   Horizontal bars on
    the  experimental  curve   represents  the  standard  deviation  and
    vertical  bars  represent  the  mean  value of  the  axial  velocity
    integrated on a $R+\delta r$  radius disc.  The tips of the vertical
    bars represent  the integration diameter  for the estimation  of the
    mean values (\textit{i.e.}  $2(R+\delta r)$).}
  \label{fig:profil_1hydrol_vit_num_exp}
\end{figure}

A closer  look can  be taken  by considering a  velocity profile  at one
particular         distance          behind         the         turbine.
Figure~\ref{fig:profil_1hydrol_vit_num_exp} shows  both experimental and
numerical profiles 1.2 diameter behind the turbine.  From these profiles
one can also estimate  the mean value of the  axial velocity $\bar{u}(x)$ at
position $x$ integrated on a $R^*=R + \delta r$ radius disc:
\begin{equation}
  \label{eq:mean_velocity}
  \bar{u}(x) = \frac{1}{{R^*}^2} \int_{-R^*}^{R^*} |y| u(x,y) \, \d y
\end{equation}
Here $\delta r  =0.05\mathrm{m} \simeq 0.14 R$ is considered,  which enlarges
the  integration interval  to the  two nearest  experimental measurement
nodes outside the rotor.  In  that manner, the whole velocity deficit is
accounted for.  The $R^*$ radius disc thus represents the turbine's area
of influence, which is  slightly larger than the turbine's cross-section
area. The  numerical profile fits  quite well with the  experiment, even
though the mean value seems  to be slightly underestimated and the shape
of the deficit is not accurately represented at the centre.  This can be
explained by the numerical  turbulence model, which is not sophisticated
enough at the present time and will be improved in the near future.

\begin{figure}[ht]
  \centering
  \includegraphics[width=0.6\linewidth]{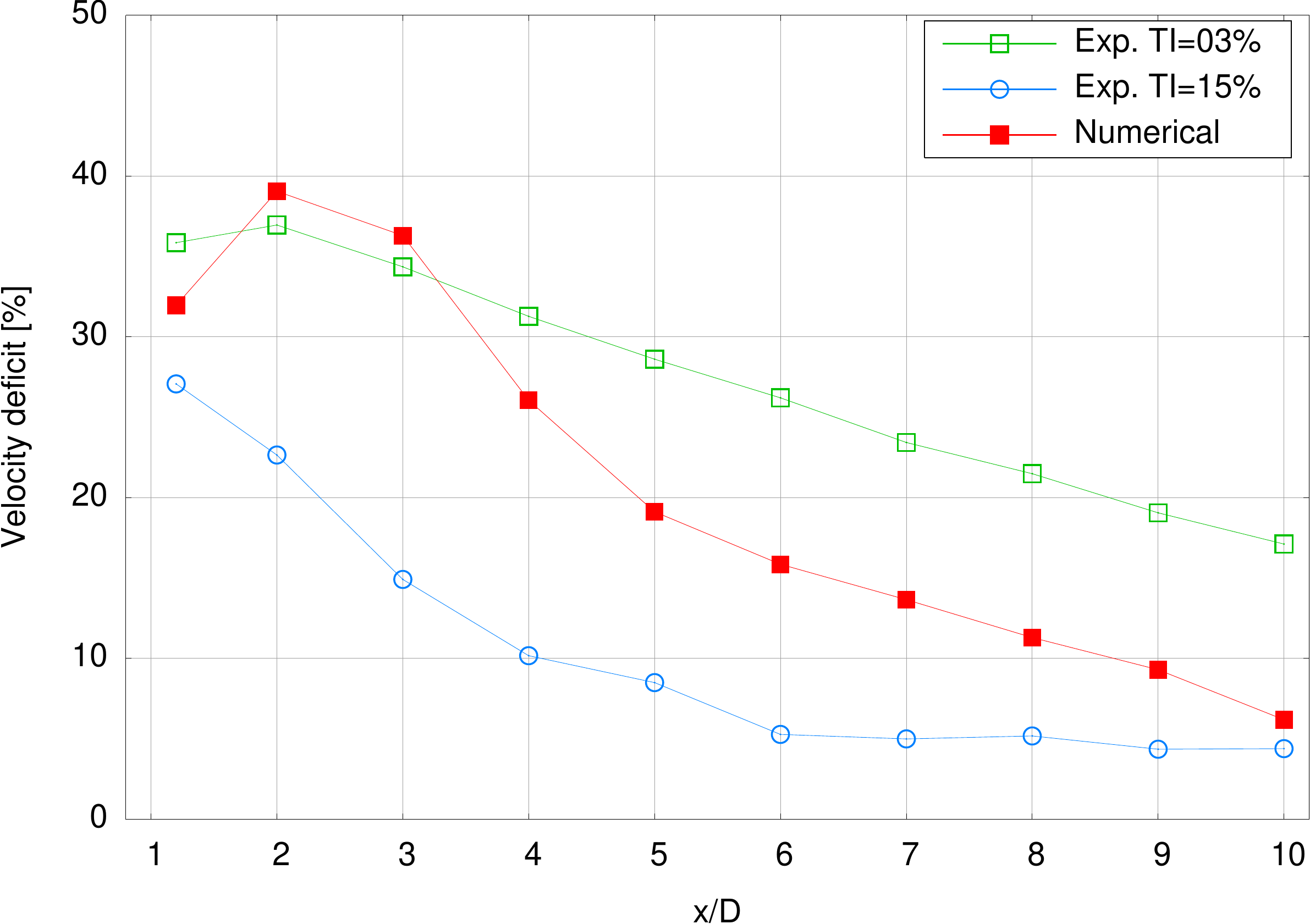}
  \caption{Mean velocity  deficit behind the  turbine for varying  TI as
    values of 3\%  and 15\% in the experiment, and a  TI of 0\% together
    with a turbulence model in the computation.}
  \label{fig:deficit_vit_1hydrol}
\end{figure}

Now  that  the  evaluation of  the  axial  velocity  mean value  on  the
turbine's  area  of influence  has  been  defined,  one can  examine  the
reduction  of the  velocity deficit  as  the distance  from the  turbine
increases.   The mean  axial  velocity  deficit $\gamma$  (in  \%) at  a
specific location $x$ behind the turbine is defined as:
\begin{equation}
  \label{eq:deficit}
  \gamma(x)=100 (1-\bar{u}(x))
\end{equation}
This definition leads to  figure~\ref{fig:deficit_vit_1hydrol}, on  which  one can see
that with an ambient 3\%  TI, the experimental velocity deficit steadily
decreases from  about 35\% at $x<2D$  down to 15\% at  $x=10D$.  On the
other hand, with  a 15\% TI, the deficit decreases  sharply in the near
wake and then  levels off around 8\% from  $x=5D$.  The relation between
the  velocity deficit  (and  therefore  the velocity  as  well) and  the
distance from the  turbine seems to be linear in the  case of an ambient
3\%  TI.   An  equivalent  numerical  curve corresponding  to  the  wake
presented in figure~\ref{fig:wake_1_hydrol_TSR3_67_num} is also shown in
order to check that the wake dynamics is numerically well reproduced.  A
better turbulence model should enable  us to compute more accurately and
to distinguish the 3\% and 15\% TI configurations.

Another  aspect of  the behaviour  of a  marine current  turbine  can be
deduced  from  the forces  and  moments  on  the turbine  blades.   More
particularly,  the axial  moment  or  torque is  used  to determine  the
turbine power coefficient $C_P$ that assesses its performance.  Besides,
the axial force can provide information about the fatigue of the machine
and is used for the calculation of the thrust coefficient $C_T$.

The power  coefficient is  defined as the  proportion of  power $\mc{P}$
retrieved by the turbine as compared to the maximum power available from
the incoming flow through the rotor area:
\begin{equation}
  \label{eq:CP}
  C_P=\frac{\mc{P}}{\frac{1}{2}\rho        S        U_\infty^3}        =
  \frac{\mc{M}_x\Phi}{\frac{1}{2}\rho    \pi     R^2    U_\infty^3}    =
  \frac{\mc{M}_x}{\frac{1}{2}\rho \pi R^3 U_\infty^2}\times TSR
\end{equation}
where $\rho$ is the density of  the fluid, $S$ is the cross-section area
of the turbine  and $\mc{M}_x$ is the axial moment,  also referred to as
the  turbine  torque,  defined  as  the  $x$-component  of  the  moment.
Similarly,  the thrust  coefficient is  defined as  the axial  force $T$
acting  upon  the turbine  as  compared to  the  kinetic  energy of  the
incoming flow through $S$:
\begin{equation}
  \label{eq:CT}
  C_T= \frac{T}{\frac{1}{2}\rho    \pi     R^2    U_\infty^2}= 
  \frac{T}{\frac{1}{2}\rho    \pi     R^3    U_\infty \Phi} \times TSR
\end{equation}

\begin{figure}[ht]
  \centering
  \subfigure[$C_P$ function of TSR\label{fig:perfo_single_cp}]{\includegraphics[width=0.48\linewidth]{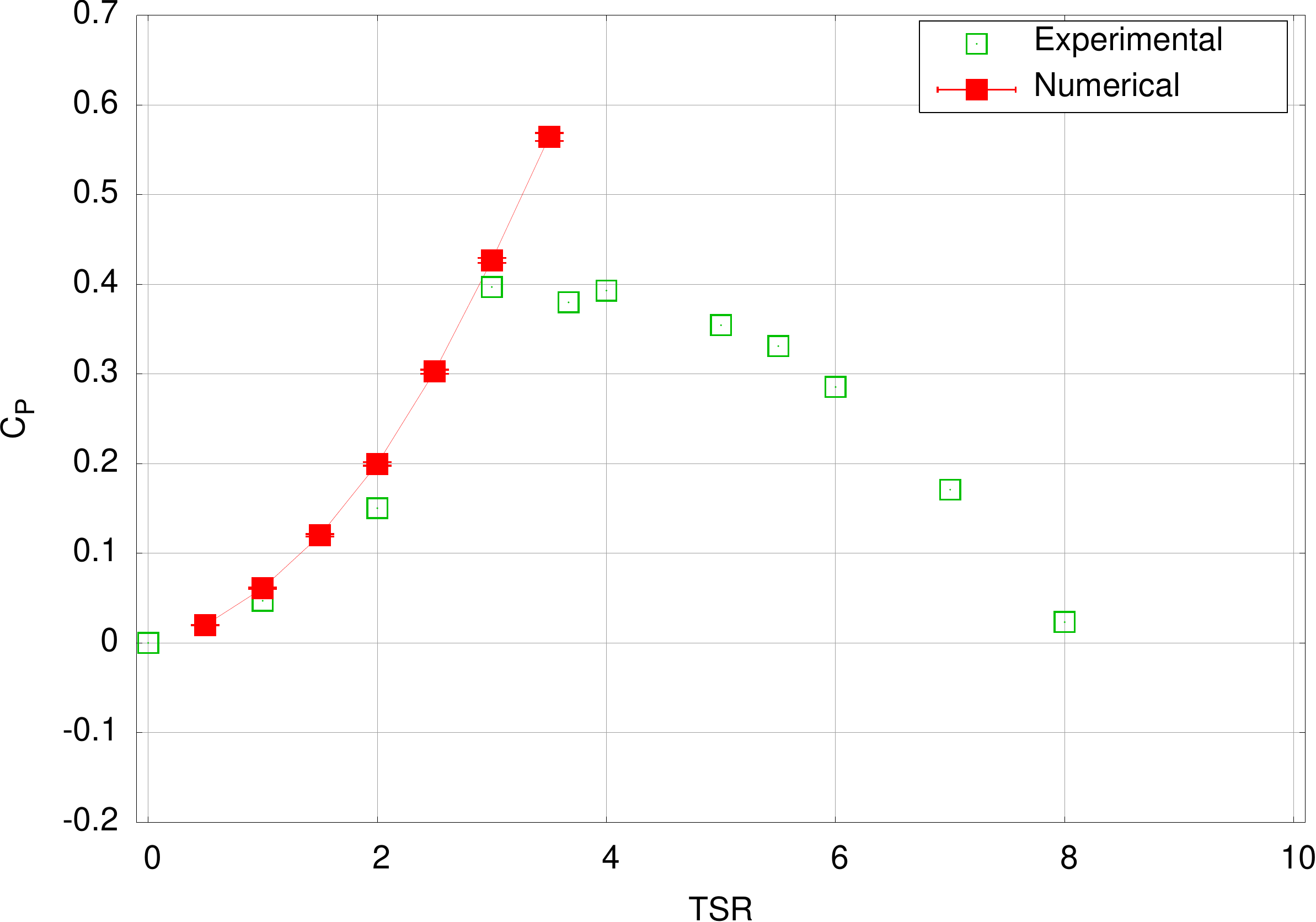}}
  \subfigure[$C_T$ function of TSR\label{fig:perfo_single_ct}]{\includegraphics[width=0.48\linewidth]{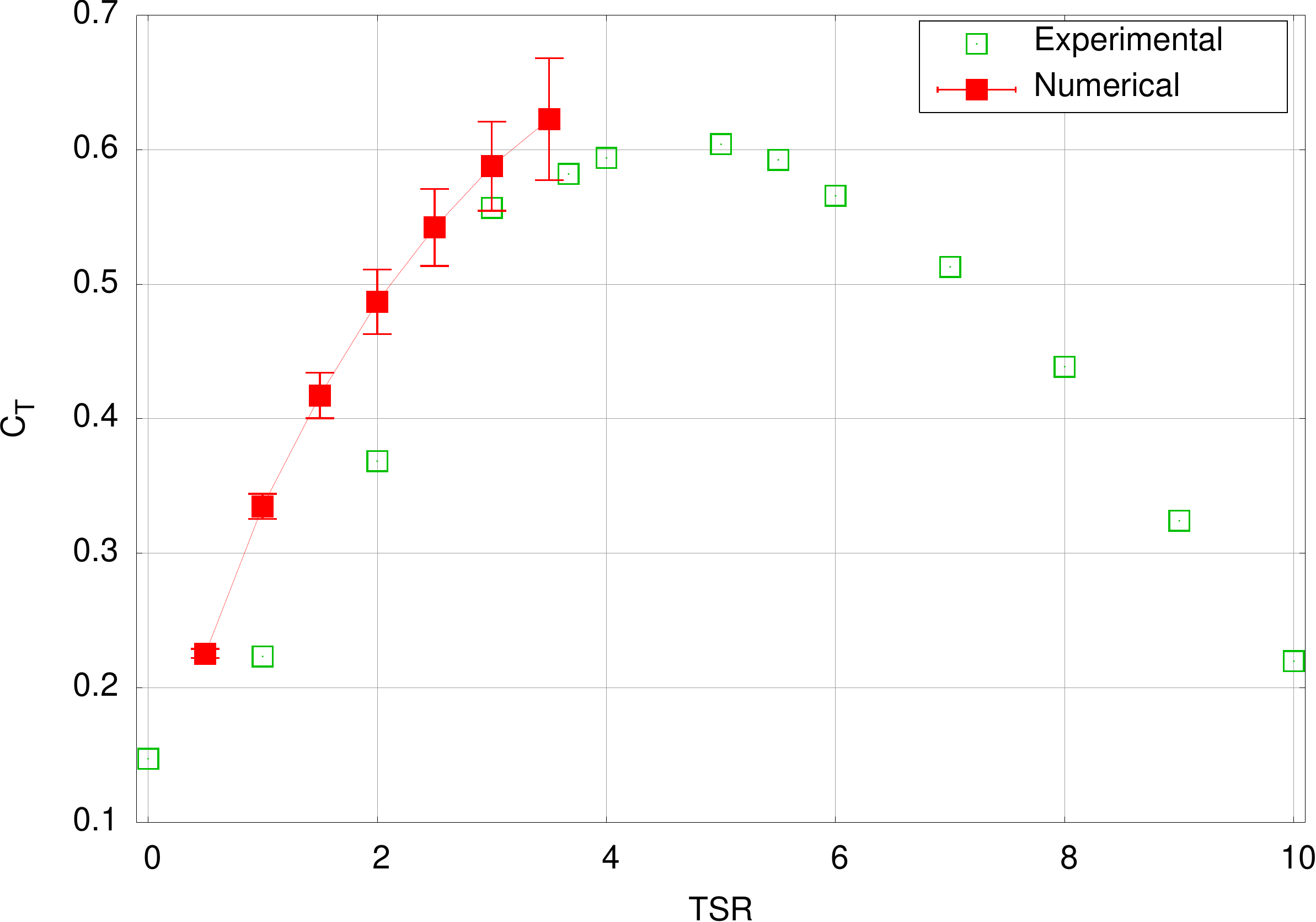}}
  \caption{Evaluation of Power  ($C_P$) and Thrust ($C_T$) coefficients,
    with  a  3\%  ambient  TI  in  the  experiment.   The  corresponding
    evolution for the computations has been  obtained with a  TI of 0\%
    with  a turbulence model.  The numerical  points represent  the mean
    values computed over $\Delta t = 2$s and the vertical bars represent
    the standard deviation with respect to this mean value.}
  \label{fig:perfo_single}
\end{figure}

For a given current velocity of 0.8m/s, and an ambient TI of 3\%,
figure~\ref{fig:perfo_single} presents the evolution of power and thrust
coefficients of a turbine function of its TSR.
Figure~\ref{fig:perfo_single_cp} shows that approximately 40\% of the
total available power is retrieved by the turbine ($C_P\simeq 0.4$) when
its TSR is between three and four. Once again, these curves are compared
to results obtained from numerical computations, which show very
good agreement.  However, the numerical particle emission model does not
account for flow separation, which explains why for TSR higher than
three, the $C_P$ keeps increasing in the numerical computations while it
should reach a peak and then decrease.  The modelling of flow separation
with a vortex method is currently being considered and should be
implemented soon.

Other configurations and blades have been tested to validate the
software in terms of $C_P$ and $C_T$, in particular
from~\cite{Bahaj2007}, to examine the influence of the blades set angle
and to check the convergence of the method~\cite{Pinon2012}.  As pointed
out, development is still necessary to improve the accuracy of 
computations.  More particularly, the turbulence model needs to be more
sophisticated, which is why one of the first priority is to find an
adequate model amongst those presented in~\cite{Meneveau2000}, for
instance.  In order to assess correctly a turbine performance, the
emission model of the numerical method also needs to be altered so as to
account for flow separation.  Once  those modifications are achieved,
focus will be made on the modelling of multi-device
configurations and corresponding validation will be possible thanks to the
experimental results presented in the next section of this paper.  The
main interest of the fully-validated software will be to model with
accuracy more complex layouts that cannot be set up in experimental
trial facilities.

%% file: montage_cote_south.pdf_tex

\begingroup
  \makeatletter
  \providecommand\color[2][]{%
    \errmessage{(Inkscape) Color is used for the text in Inkscape, but the package 'color.sty' is not loaded}
    \renewcommand\color[2][]{}%
  }
  \providecommand\transparent[1]{%
    \errmessage{(Inkscape) Transparency is used (non-zero) for the text in Inkscape, but the package 'transparent.sty' is not loaded}
    \renewcommand\transparent[1]{}%
  }
  \providecommand\rotatebox[2]{#2}
  \ifx\svgwidth\undefined
    \setlength{\unitlength}{514.36040039pt}
  \else
    \setlength{\unitlength}{\svgwidth}
  \fi
  \global\let\svgwidth\undefined
  \makeatother
  \begin{picture}(1,0.69842502)%
    \put(0,0){\includegraphics[width=\unitlength]{montage_cote_south.pdf}}%
    \put(0.87691109,0.57715466){\color[rgb]{0,0,0}\makebox(0,0)[rb]{\smash{$x$}}}%
    \put(0.23722711,0.46048049){\color[rgb]{0,0,0}\makebox(0,0)[b]{\smash{$U_\infty$}}}%
    \put(0.37067838,0.39251484){\color[rgb]{0,0,0}\makebox(0,0)[b]{\smash{TI}}}%
    \put(0.85815619,0.44512218){\color[rgb]{0,0,0}\makebox(0,0)[lb]{\smash{$h=1.10$m}}}%
    \put(0.09558291,0.36217936){\color[rgb]{0,0,0}\makebox(0,0)[rb]{\smash{$H=2$m}}}%
    \put(0.59563877,0.00320027){\color[rgb]{0,0,0}\makebox(0,0)[b]{\smash{$L=0.7$m}}}%
    \put(0.60515942,0.04976856){\color[rgb]{0,0,0}\makebox(0,0)[b]{\smash{$l=0.6$m}}}%
    \put(0.88324851,0.34249985){\color[rgb]{0,0,0}\makebox(0,0)[lb]{\smash{$D=0.7$m}}}%
    \put(0.53154726,0.33962528){\color[rgb]{0,0,0}\makebox(0,0)[lb]{\smash{$O(0;0;0)$}}}%
    \put(0.17094657,0.68424478){\color[rgb]{0,0,0}\makebox(0,0)[rb]{\smash{$e_z$}}}%
    \put(0.22945561,0.6213809){\color[rgb]{0,0,0}\makebox(0,0)[lb]{\smash{$e_x$}}}%
    \put(0.17701521,0.581528){\color[rgb]{0,0,0}\makebox(0,0)[rb]{\smash{$e_y$}}}%
  \end{picture}%
\endgroup

%% file: array.tex
\section{Marine current turbines interaction}

\subsection{General considerations}
\label{sec:array_general}

Studies  concerning the  layout of  marine current  turbines  arrays are
still   few.   However,   general  guidelines   and   \textit{a  priori}
considerations  can  be   found  in  recent  literature~\cite{Myers2010,
  Rawlinson-Smith2010}.  Two  kinds of arrays have  to be distinguished,
first generation and second generation  arrays, so called because of the
probable progressive row by row  growth of these farms. First generation
arrays designate  the youngest farms, made  up of a single  or two rows,
designed to avoid  any interaction effect between the  turbines.  On the
contrary, second generation  arrays refer to larger arrays in
which such interactions cannot be avoided~\cite{Myers2010}.

\begin{figure}[ht]
  \centering
  \resizebox{0.7\linewidth}{!}{\Large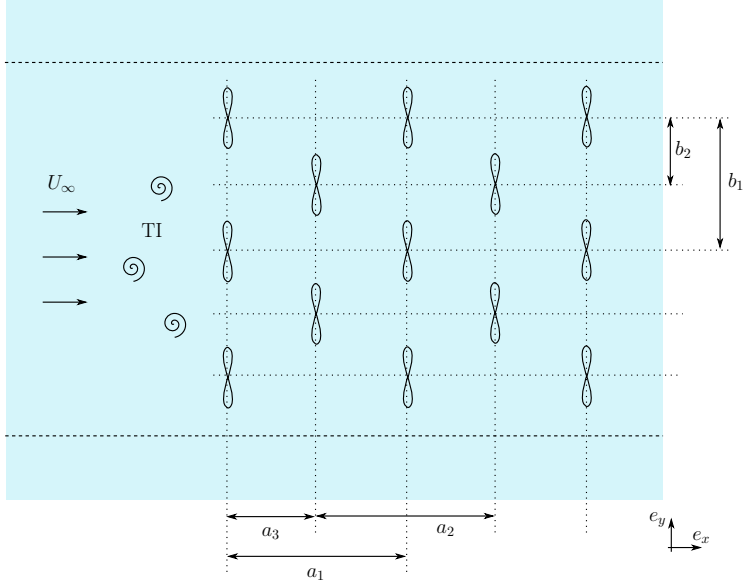}
  \caption{Schematic top view of a marine current turbines array.}
  \label{fig:schem_array}
\end{figure}

It  is then  clear  that  those two  designations  have a  chronological
meaning. As  a matter of  fact, at an  early age of  their implantation,
arrays will be  made up of one  or two rows of turbines,  the second row
being placed downstream  the first one and shifted so  that the wakes of
the upstream devices  will not interact with the  turbines of the second
row. The early extensions of such arrays will be performed by adding new
devices into the one or two existing rows, but a time will come when the
only way to enlarge first generation arrays will consist in adding rows.
From  then   on,  interactions  between  turbines  will   no  longer  be
avoidable. Figure~\ref{fig:schem_array} illustrates such a layout.

Several parameters, besides  the depth of the array in  the flow that is
considered here to be far enough from both  the free surface and the sea bed in
order  to neglect  their  interaction, are  characteristic  of an  array
layout: 
\begin{itemize}
  \item The distance $a_1$ between two successive ``even'' rows;
  \item The distance $a_2$ between two successive ``odd'' rows;
  \item The  distance $a_3$  between an upstream  even row and  the very
    next (odd) row;
  \item The distance $b_1$ between two adjacent turbines of a same row;
  \item   The   $y$-offset   $b_2$   between  two   successive   current
    perpendicular lines of turbines.
\end{itemize} 

Some suggestions can be made concerning those parameters. In particular,
it would  be natural  to consider that  $b_2=\frac{1}{2}b_1$, as  it has
been suggested  in~\cite{Myers2010}. Similarly, parameters  $a_1$, $a_2$
and $a_3$ can  be chosen such that $a_1=a_2=2a_3$.   However, the choice
of a smaller $a_3$ would also make sense so as to benefit from potential
``positive'' interactions from the upstream turbines.

The present study focuses  on the layout of second  generation arrays issue, and
more  particularly on  the characterization  of the  interaction effects
between  two marine current  turbines placed  one behind  another. Hence
parameter $a_1$ alone describes  the configuration, and will then simply
be referred to as $a$.

\subsection{Experimental setup}
\label{sec:exp_setup}

\begin{figure}[ht]
  \centering
  \includegraphics[width=0.7\linewidth]{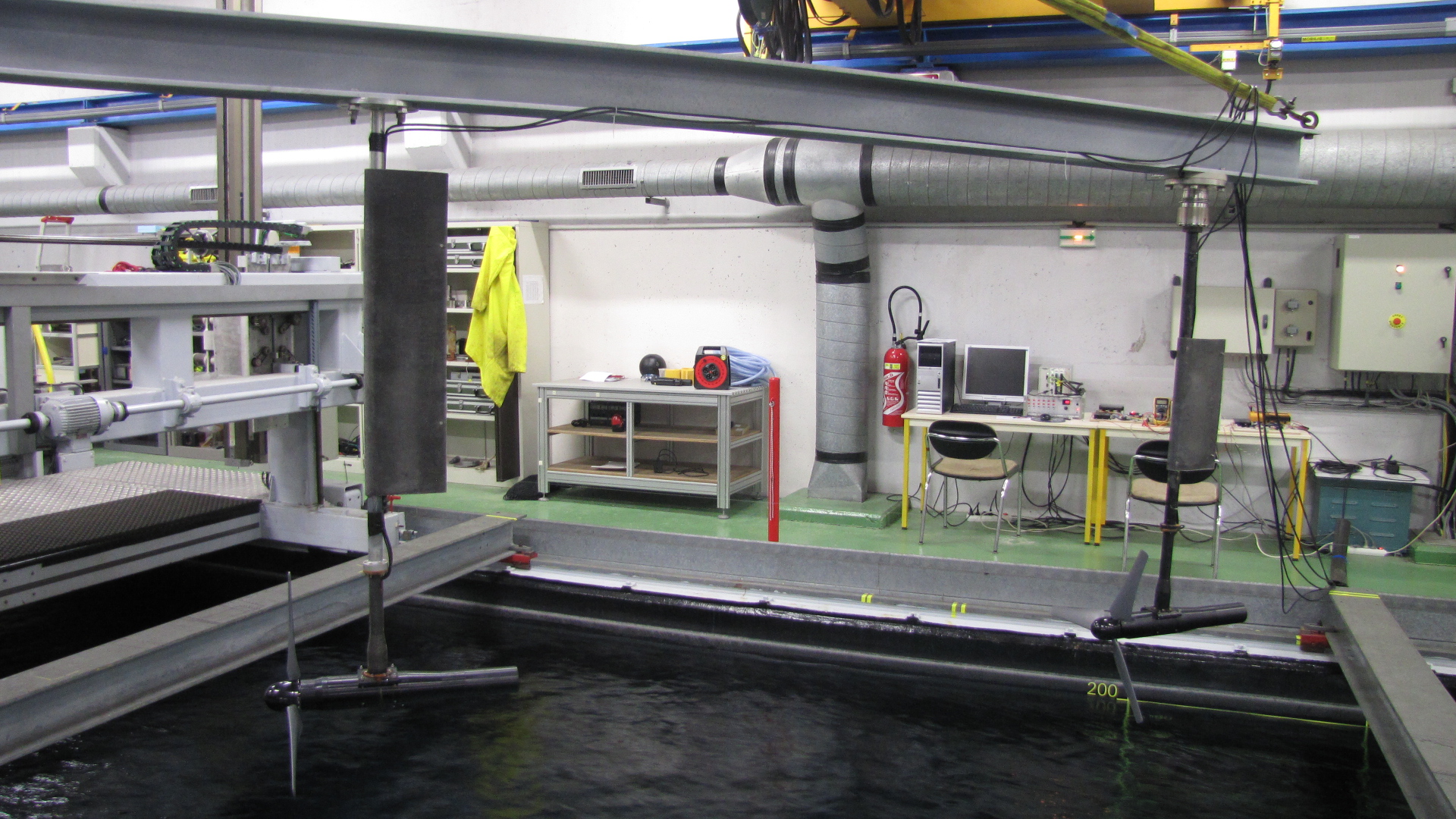}
  \caption{Photography of the experimental  setup with $a=4D$.  For $a >
    5D$ configurations, the lengthwise  girder is split into two smaller
    girders  and two additional  transerse girders  are used  to support
    them (cf. figure \ref{fig:schem_setup}).}
  \label{fig:photo_setup}
\end{figure}

The  experimental setup  is made  up  of two  1/30\up{th} scale  turbine
models  attached to  one (for  $a\leq  5D$ configurations)  or two  (for
$a>5D$ configurations)  lengthwise girder(s) thanks to two  poles of the
same  length.  A  photography of  the  setup  with  $a=4D$ is  given  in
figure~\ref{fig:photo_setup}.

\begin{figure}[!ht]
  \centering
  \subfigure[Side view]{\resizebox{0.7\linewidth}{!}{\Large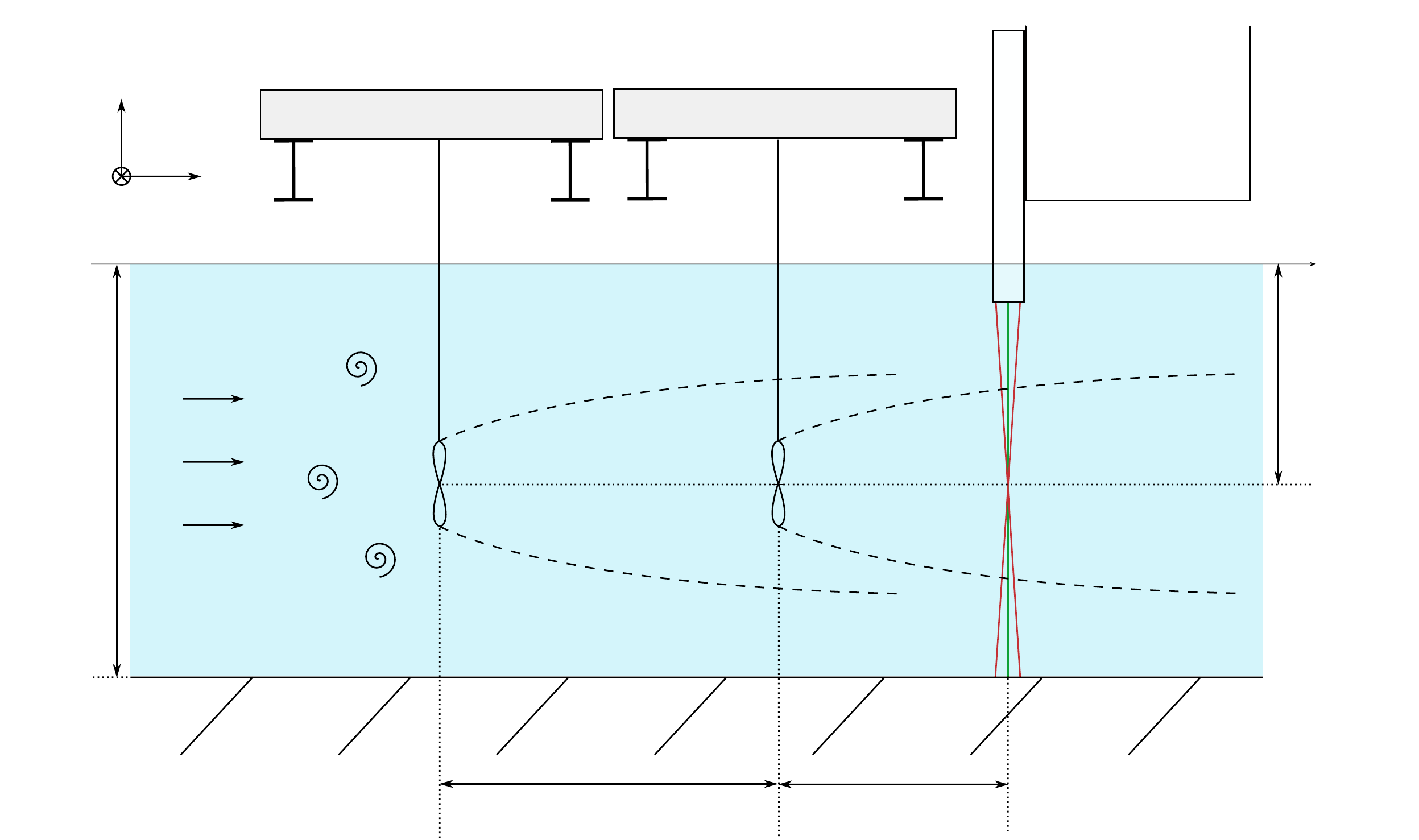}}\\
  \subfigure[Top view]{\resizebox{0.6\linewidth}{!}{\Large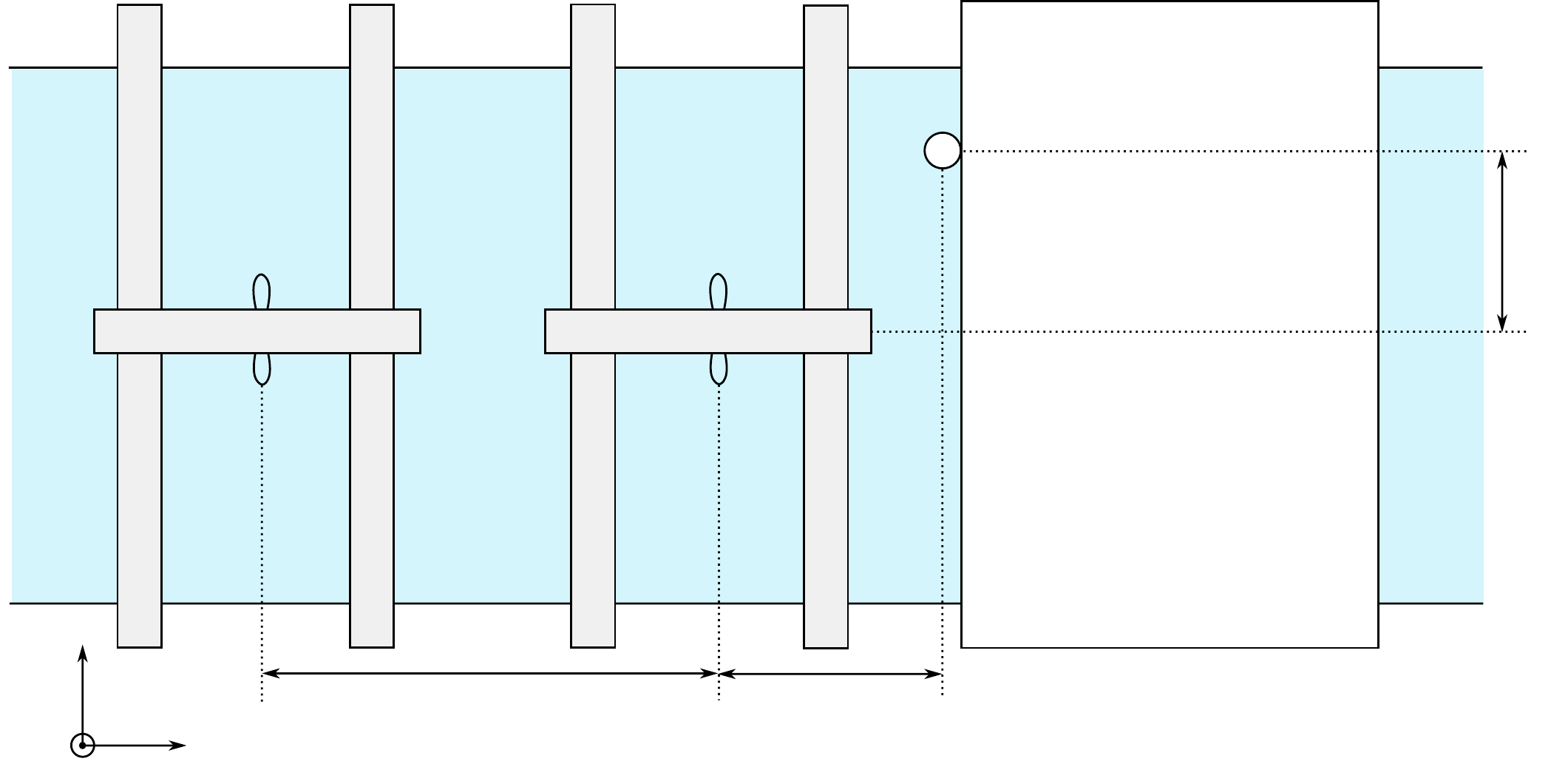}}
  \caption{Schematic view  of the experimental  setup for configurations
    with $a>5D$.  The origin $O(0;0;0)$ is  chosen at the  center of the
    downstream turbine rotor.}
  \label{fig:schem_setup}
\end{figure}

The  girder is  placed over  the flume  tank, parallel  to  the upstream
current and at equal distance from the two sides. The downstream turbine
is equipped with a six-component  load cell and a two-component laser is
attached to  a footbridge over the  flume. The laser can  move along the
footbridge that  can itself  be shifted backward  and forward  along the
flume. The Laser Doppler Velocimetry  (LDV) technique is used to measure
axial and radial velocities at  different locations on a grid behind the
downstream  turbine. This  allows to  draw  maps such  as those  of
figure~\ref{fig:wake_1hydrol_TSR3_67}              shown              in
section~\ref{sec:single_wake}. Schematic views of the complete setup are
given in figure~\ref{fig:schem_setup}.

\subsection{Wake interactions}
\label{sec:array_wake}

This study of wake interactions focuses on  a $a=4D$ configuration, with both  TSR=3.67 and an upstream
3\% ambient TI. This is motivated  by the results obtained with a single
turbine (cf.  section~\ref{sec:single_wake}). As  a matter of fact, when
looking at figure~\ref{fig:wake_1hydrol_TSR3_67_turb5}, one can see that
the turbulence  intensity rate four  diameters behind the turbine  is in
the  order of 15\%.   It should thus be relevant  to compare  this case  to the
single  device configuration in  a 15\%  ambient TI  due to  the current
generation  process   in  the  flume  tank   without  any  ``smoothing''
techniques (\textit{e.g.}   using a honeycomb to reduce  the ambient TI;
for more  detail, see~\cite{Maganga2010}).  If the  downstream device in
the $a=4D$ configuration  behaves as if it were  single, this would mean
that there is no wake  interactions \textit{per se} and that the ambient
TI  rate represents  the  only affecting  ``input''  parameters for  the
behaviour of interacting turbines.

\begin{figure}[ht]
  \centering
  \subfigure[Axial velocity map]{\includegraphics[width=0.48\linewidth]{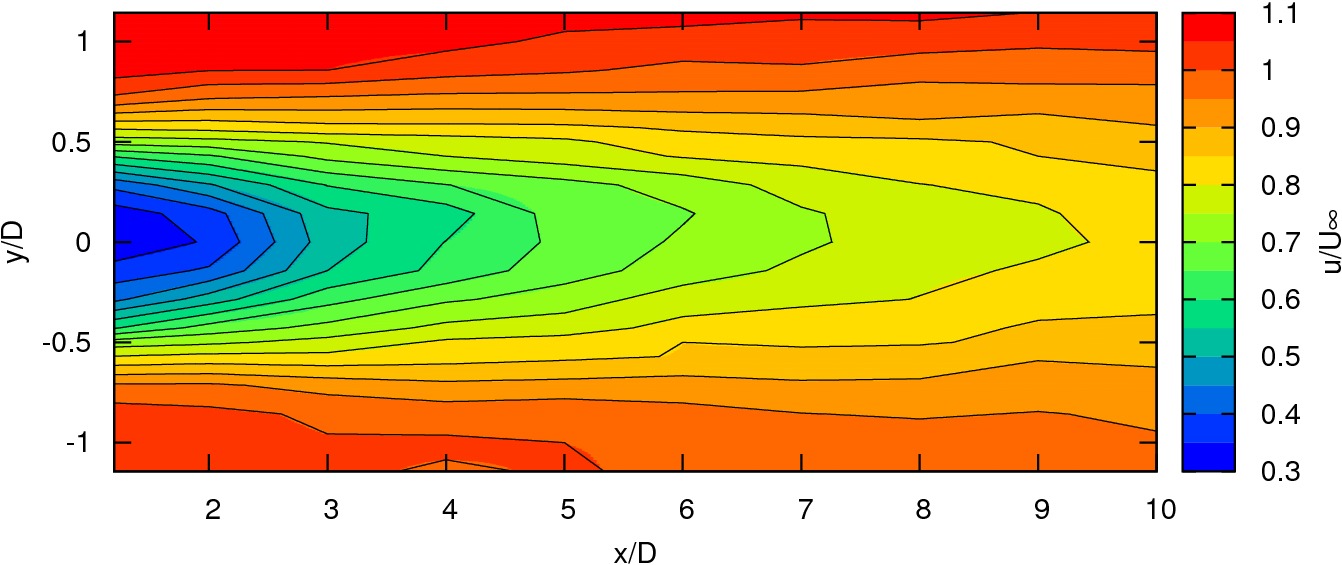}}
  \subfigure[Turbulence map]{\includegraphics[width=0.48\linewidth]{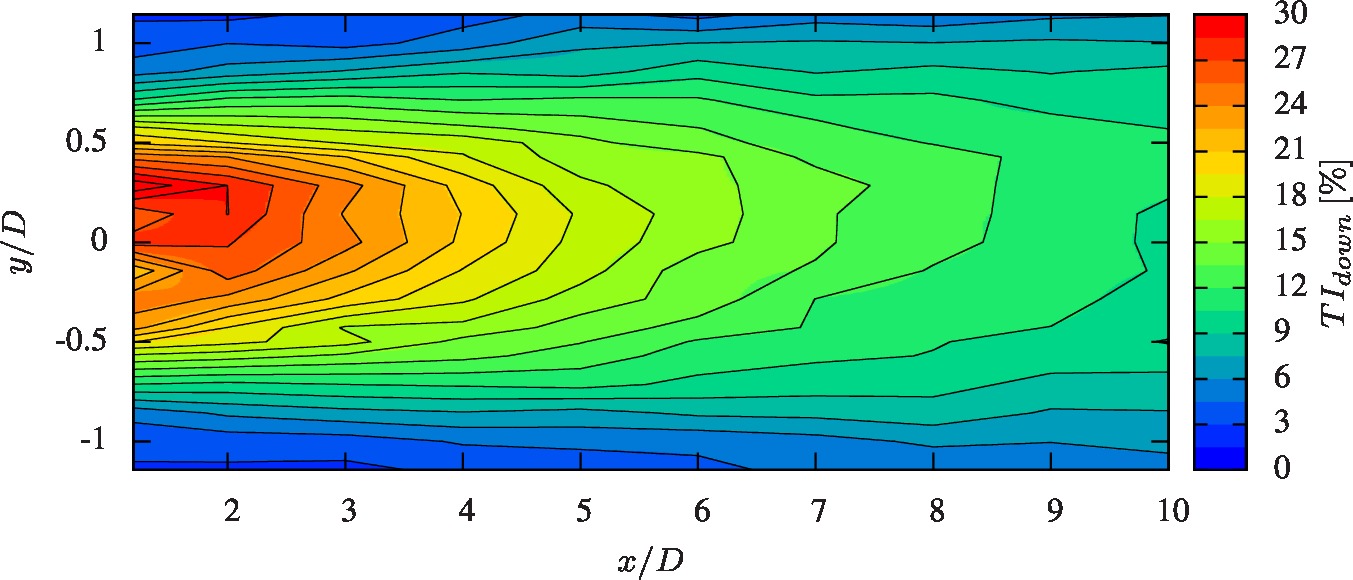}}
  \caption{Wake behind  the second turbine with  $a=4D$, upstream TI=3\%
    and both TSR=3.67.}
  \label{fig:wake_2hydrol_TSR3_67}
\end{figure}

In          the           single          device          configuration,
figures~\ref{fig:wake_1hydrol_TSR3_67_vit25}
and~\ref{fig:deficit_vit_1hydrol} show that  the velocity deficit behind
a  turbine   in  a  15\%   turbulent  upstream  flow  tends   to  reduce
rapidly. This represents a significant  advantage with a view to place a
second turbine in the  wake of the first one. On the  other hand, in the
case of interacting turbines,  axial velocity and turbulence maps behind
the         downstream        turbine        are         drawn        in
figure~\ref{fig:wake_2hydrol_TSR3_67} .  Since the downstream turbine is
placed in an  area where the mean turbulence intensity  rate is close to
15\%, the general aspect of its wake  may look like the wake of a single
turbine in  a 15\%  ambient TI. Unfortunately,  one can easily  see that
this is not  the case, which indicates that there  is a real interaction
between the two turbines.

This  suggests that  the turbulence  induced by  the  current generation
process  and the  one induced  by the  turbine do  not present  the same
characteristics.   A complementary  study was recently  carried out
with two  turbines in a 15\%  upstream ambient TI in  order to determine
whether  the presence of  the upstream  turbine modifies  the turbulence
structure~\cite{Mycek2013b}.

\subsection{Downstream performance}
\label{sec:down_perfo}

\begin{figure}[!ht]
  \centering
  \subfigure[$C_P^{down}$ of  the downstream  device function of  its TSR,
    with an upstream TSR of 3 and  an ambient TI of 3\%, compared to the
    $C_P^{single}$    of   a   single    turbine   shown    in   section
    \ref{sec:single_comp_num}.\label{fig:cp_twin}]{\includegraphics[width=0.48\linewidth]{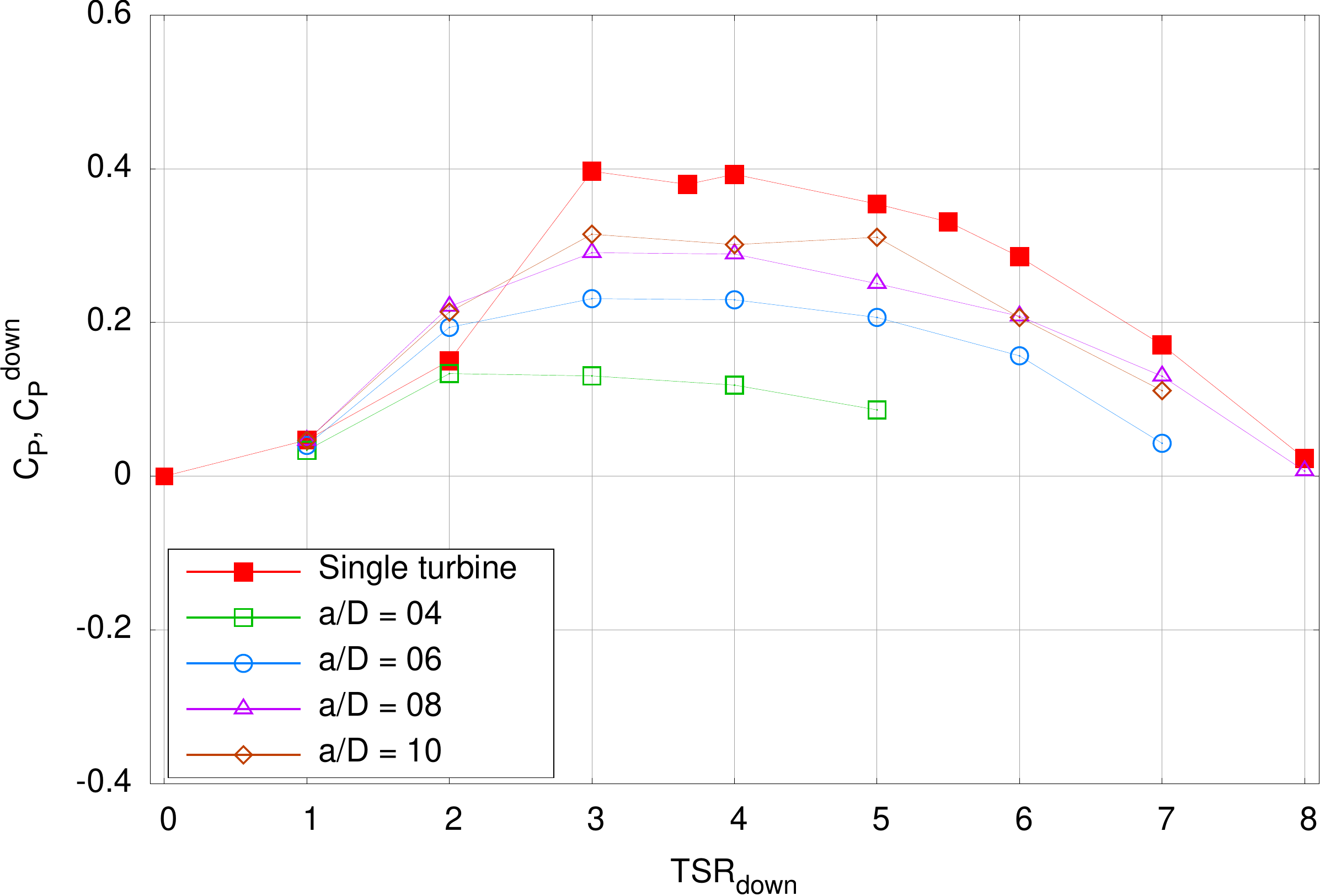}}\hfill
    \subfigure[Maximum  $C_P^{down}$  obtained  on  the  downstream  turbine
    function of  the inter-device  distance, for two  different upstream
    TSR.\label{fig:cp_max}]{\includegraphics[width=0.48\linewidth]{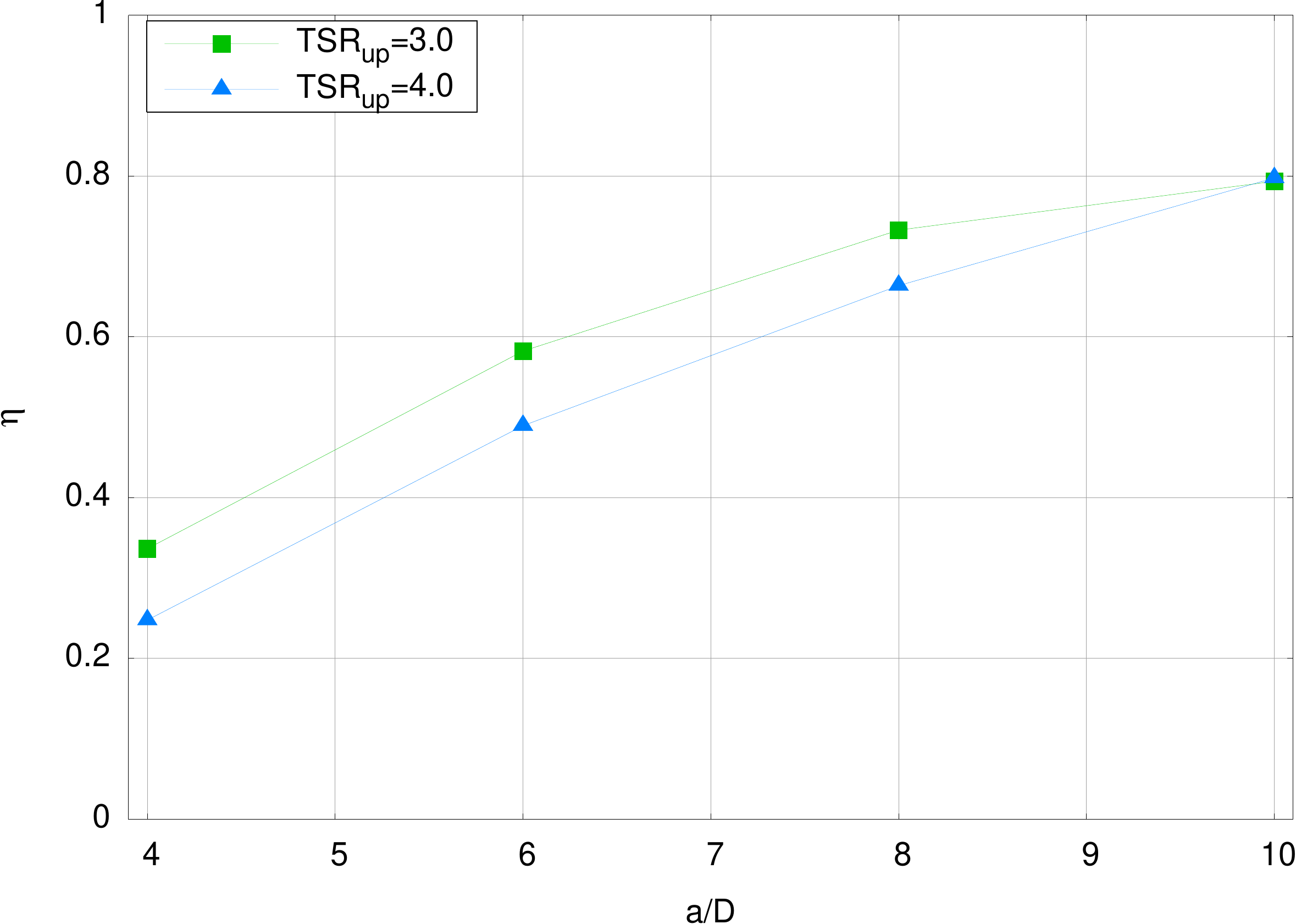}}
    \caption{Performance evaluation of the twin-device configuration.}
\end{figure}

Another way to evaluate the  interaction effects between two turbines is
to examine  the performance of  the downstream turbine.   The comparison
with the  performance of a single device  gives an idea on  how deeply a
turbine is affected by the presence of an upstream device. The evolution
of the  downstream turbine  $C_P^{down}$ is plotted  against its  TSR on
figure~\ref{fig:cp_twin}, for  different $a/D$ configurations,  and with
an upstream device TSR of three, which yields maximum individual energy.
The evolution of  the single device $C_P^{single}$ function  of its TSR,
already shown in section \ref{sec:single_comp_num}, is also plotted as a
matter of comparison. It should be  noted that the TSR of the downstream
turbine is  computed thanks to  equation~\eqref{eq:TSR} where $U_\infty$
is the upstream velocity before  the first device, and thus \textit{not}
the mean velocity  at the location of the second  device. This choice is
motivated by the  fact that in real conditions, one will  not be able to
access  the   actual  velocity  at   the  location  of   the  downstream
turbine.  The present study  of  interaction effects  between  turbines is  thus
carried  out  considering only  ``measurable''  upstream quantities.  By
doing so,  general conclusions  may be drawn about the behaviour of
interacting   turbines  depending  only   on  those   input  parameters.
Similarly, the  downstream turbine  $C_P^{down}$ is still  computed from
the upstream  velocity $U_\infty$.  It  is then important  to understand
that $C_P^{down}$ is  an abuse of notation since  this quantity does not
represent  any power coefficient;  it can  only be  an indicator  of the
power retrieved as compared to the upstream velocity $U_\infty$.

The experiment  shows that the  maximum $C_P^{down}$ for  the downstream
turbine  is obtained  with a  downstream turbine  TSR between  three and
four,  except for  the $a=4D$  configuration where  the  $C_P^{down}$ is
slightly better  with TSR=2.   One can also  notice that the  longer the
distance between  the two  devices is, the  better the evolution  of the
downstream   turbine   $C_p^{down}$   fits   with   a   single   turbine
$C_P^{single}$.   This is  clearly due  to  the fact  that the  velocity
deficit generated by  a turbine decreases as a  function of the distance
from the device (cf. figure~\ref{fig:deficit_vit_1hydrol}).

This  trend  can be  confirmed  by plotting  the  value  of the  maximum
$C_P^{down}$ obtained for each  $a/D$ configuration. The ``efficiency'' $\eta$ is defined as
the ratio  of the maximum $C_P^{down}$  of the downstream  device to the
maximum  $C_P^{up}$ of  the  upstream device,  that  is to  say the  one
obtained with TSR=3 for a single device:
\begin{equation}
  \label{eq:CPmax}
  \eta                                                               =
  \frac{\max\big(C_{P}^{down}(TSR)\big)}{\max\big(C_{P}^{up}(TSR)\big)}
  = \frac{\max\big(C_{P}^{down}(TSR)\big)}{C_{P}^{single}(TSR =3)}
\end{equation}
Figure~\ref{fig:cp_max} shows  the evolution  of $\eta$ function  of the
distance $a/D$,  for upstream device  TSR of 3  and 4. As  expected, the
maximum $C_P^{down}$  raises as  $a$ increases and  reaches 80\%  of the
maximum retrievable power for a  single device when $a/D=10$.  It should
be pointed out  that the downstream device can  retrieve more power when
the  upstream  turbine has  its  TSR=3  rather  than 4.   These  results
indicate   that  a   compromise  between   individual   performance  and
inter-device spacing is necessary.   Considering an implantation area of
given  shape  and surface,  the  more  distant  two successive  rows  of
turbines, the higher  the individual power retrieved; but  there is then
less space  for additional rows  of devices, that  is to say  that fewer
turbines can  be implanted. Hence a  complete compromise has  to be made
considering the implantation of marine current turbines farms.

\subsection{Comparison with numerical results}
\label{sec:array_comp_num}

Coarse numerical simulations were run with two turbines with $a=4D$, and
$TSR=TSR^{up}=TSR^{down}=3.67$. Since there is no maximum in the
numerical $C_P$ curves, the ratio $r_{C_P}$ for two turbines
with the same TSR is defined as follows:
\begin{equation}
  \label{eq:ratioCP}
  r_{C_P}(TSR)=\frac{C_P^{down}(TSR)}{C_P^{single}(TSR)}
\end{equation}
Similarly, $r_{C_T}$ is defined by:
\begin{equation}
  \label{eq:ratioCT}
  r_{C_T}(TSR)=\frac{C_T^{down}(TSR)}{C_T^{single}(TSR)}
\end{equation}

\begin{figure}[!ht]
  \centering
  \subfigure[$r_{C_P}$]{\includegraphics[width=0.48\linewidth]{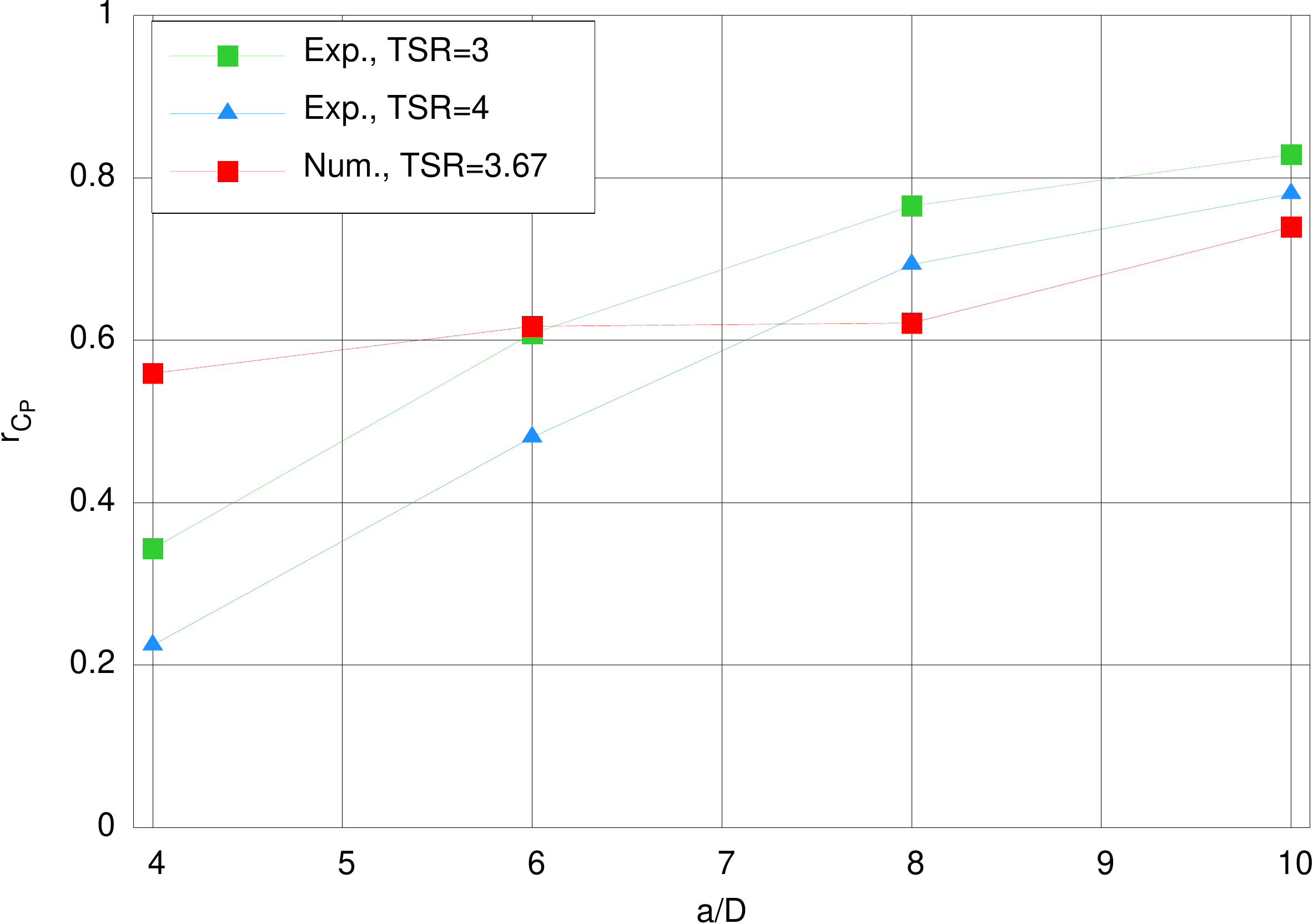}}\hfill
    \subfigure[$r_{C_T}$]{\includegraphics[width=0.48\linewidth]{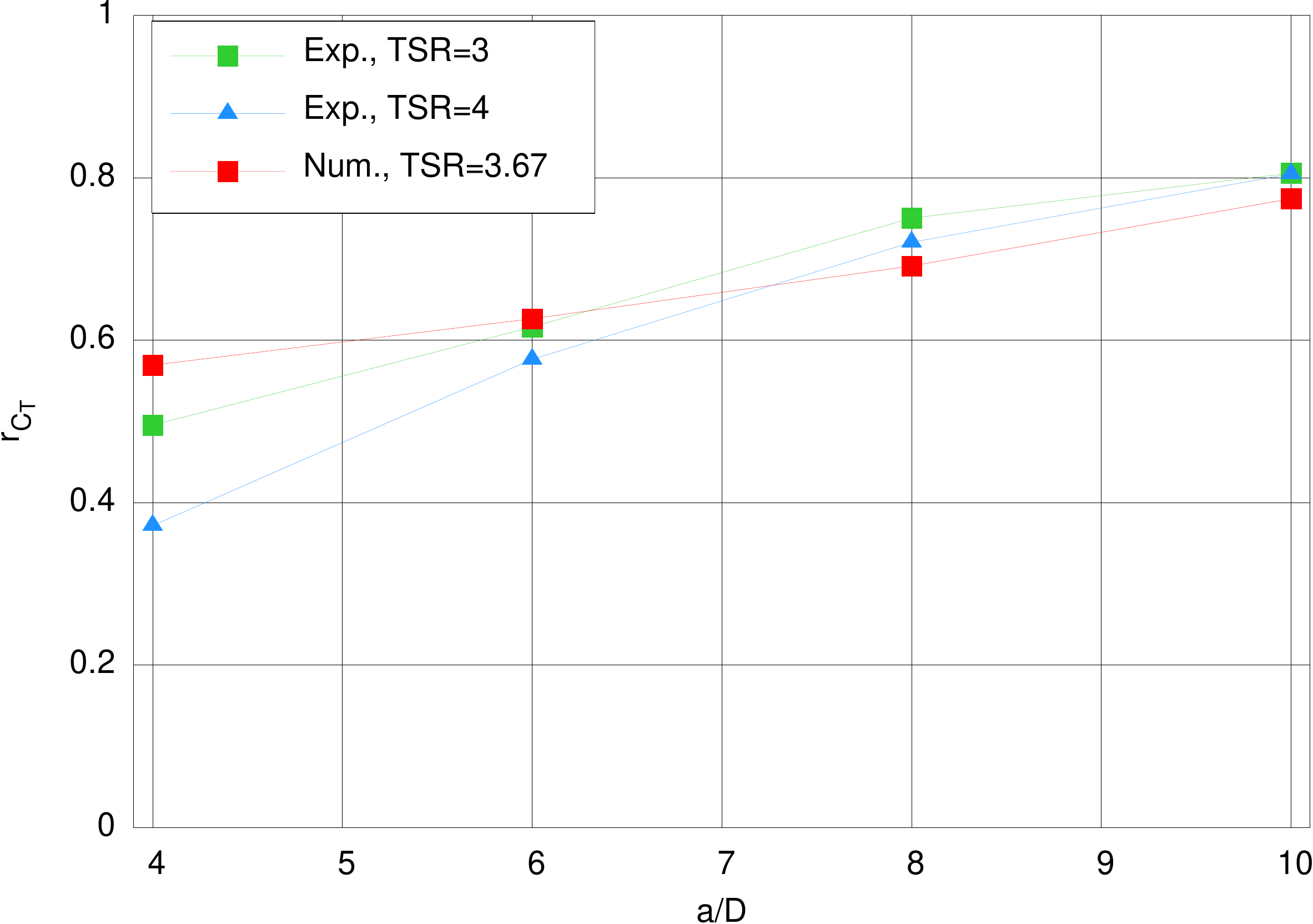}}
    \caption{Numerical comparison of $C_P$ and $C_T$ ratios for a given $TSR=TSR^{up}=TSR^{down}$.\label{fig:ratios}}
\end{figure}

Figure~\ref{fig:ratios} depicts the evolution of these ratios $r_{C_P}$
and $r_{C_T}$ function of the inter-device distance $a$, for two
experimental configurations and one numerical configuration. The
comparison between numerical and experimental results show a good
agreement, even if the discretisation is coarse and the results are not
converged. This is promising for a future realistic turbine array
numerical modelling.


%% file: ferme_hydro.pdf_tex

\begingroup
  \makeatletter
  \providecommand\color[2][]{%
    \errmessage{(Inkscape) Color is used for the text in Inkscape, but the package 'color.sty' is not loaded}
    \renewcommand\color[2][]{}%
  }
  \providecommand\transparent[1]{%
    \errmessage{(Inkscape) Transparency is used (non-zero) for the text in Inkscape, but the package 'transparent.sty' is not loaded}
    \renewcommand\transparent[1]{}%
  }
  \providecommand\rotatebox[2]{#2}
  \ifx\svgwidth\undefined
    \setlength{\unitlength}{545.39477539pt}
  \else
    \setlength{\unitlength}{\svgwidth}
  \fi
  \global\let\svgwidth\undefined
  \makeatother
  \begin{picture}(1,0.75440261)%
    \put(0,0){\includegraphics[width=\unitlength]{ferme_hydro.pdf}}%
    \put(0.07376476,0.50817769){\color[rgb]{0,0,0}\makebox(0,0)[b]{\smash{$U_\infty$}}}%
    \put(0.19116588,0.44656535){\color[rgb]{0,0,0}\makebox(0,0)[b]{\smash{TI}}}%
    \put(0.93918539,0.51099758){\color[rgb]{0,0,0}\makebox(0,0)[lb]{\smash{$b_1$}}}%
    \put(0.87065123,0.5526163){\color[rgb]{0,0,0}\makebox(0,0)[lb]{\smash{$b_2$}}}%
    \put(0.34556691,0.05832394){\color[rgb]{0,0,0}\makebox(0,0)[b]{\smash{$a_3$}}}%
    \put(0.57245522,0.06156521){\color[rgb]{0,0,0}\makebox(0,0)[b]{\smash{$a_2$}}}%
    \put(0.4039096,0.00484311){\color[rgb]{0,0,0}\makebox(0,0)[b]{\smash{$a_1$}}}%
    \put(0.85915618,0.07753851){\color[rgb]{0,0,0}\makebox(0,0)[rb]{\smash{$e_y$}}}%
    \put(0.89281392,0.05222016){\color[rgb]{0,0,0}\makebox(0,0)[lb]{\smash{$e_x$}}}%
  \end{picture}%
\endgroup

%% file: montage_cote2.pdf_tex

\begingroup
  \makeatletter
  \providecommand\color[2][]{%
    \errmessage{(Inkscape) Color is used for the text in Inkscape, but the package 'color.sty' is not loaded}
    \renewcommand\color[2][]{}%
  }
  \providecommand\transparent[1]{%
    \errmessage{(Inkscape) Transparency is used (non-zero) for the text in Inkscape, but the package 'transparent.sty' is not loaded}
    \renewcommand\transparent[1]{}%
  }
  \providecommand\rotatebox[2]{#2}
  \ifx\svgwidth\undefined
    \setlength{\unitlength}{718.27578125pt}
  \else
    \setlength{\unitlength}{\svgwidth}
  \fi
  \global\let\svgwidth\undefined
  \makeatother
  \begin{picture}(1,0.59221774)%
    \put(0,0){\includegraphics[width=\unitlength]{montage_cote2.pdf}}%
    \put(0.92533865,0.42455572){\color[rgb]{0,0,0}\makebox(0,0)[rb]{\smash{$x$}}}%
    \put(0.1476038,0.3332084){\color[rgb]{0,0,0}\makebox(0,0)[b]{\smash{$U_\infty$}}}%
    \put(0.24539639,0.2845379){\color[rgb]{0,0,0}\makebox(0,0)[b]{\smash{TI}}}%
    \put(0.43533288,0.01700017){\color[rgb]{0,0,0}\makebox(0,0)[b]{\smash{$a$}}}%
    \put(0.62655469,0.01661731){\color[rgb]{0,0,0}\makebox(0,0)[b]{\smash{$x$}}}%
    \put(0.90856685,0.33112047){\color[rgb]{0,0,0}\makebox(0,0)[lb]{\smash{$h=1.10$m}}}%
    \put(0.06844733,0.26281453){\color[rgb]{0,0,0}\makebox(0,0)[rb]{\smash{$H=2$m}}}%
    \put(0.71022729,0.58085045){\color[rgb]{0,0,0}\makebox(0,0)[b]{\smash{LDV}}}%
    \put(0.55929258,0.25729962){\color[rgb]{0,0,0}\makebox(0,0)[lb]{\smash{$O(0;0;0)$}}}%
    \put(0.07701727,0.52156181){\color[rgb]{0,0,0}\makebox(0,0)[rb]{\smash{$e_z$}}}%
    \put(0.11891584,0.47654471){\color[rgb]{0,0,0}\makebox(0,0)[lb]{\smash{$e_x$}}}%
    \put(0.08136305,0.44800588){\color[rgb]{0,0,0}\makebox(0,0)[rb]{\smash{$e_y$}}}%
  \end{picture}%
\endgroup

%% file: montage_dessus2.pdf_tex

\begingroup
  \makeatletter
  \providecommand\color[2][]{%
    \errmessage{(Inkscape) Color is used for the text in Inkscape, but the package 'color.sty' is not loaded}
    \renewcommand\color[2][]{}%
  }
  \providecommand\transparent[1]{%
    \errmessage{(Inkscape) Transparency is used (non-zero) for the text in Inkscape, but the package 'transparent.sty' is not loaded}
    \renewcommand\transparent[1]{}%
  }
  \providecommand\rotatebox[2]{#2}
  \ifx\svgwidth\undefined
    \setlength{\unitlength}{610.78969591pt}
  \else
    \setlength{\unitlength}{\svgwidth}
  \fi
  \global\let\svgwidth\undefined
  \makeatother
  \begin{picture}(1,0.49969004)%
    \put(0,0){\includegraphics[width=\unitlength]{montage_dessus2.pdf}}%
    \put(0.31671509,0.04391321){\color[rgb]{0,0,0}\makebox(0,0)[b]{\smash{$a$}}}%
    \put(0.53110963,0.04346297){\color[rgb]{0,0,0}\makebox(0,0)[b]{\smash{$x$}}}%
    \put(0.96732841,0.34709701){\color[rgb]{0,0,0}\makebox(0,0)[lb]{\smash{$y$}}}%
    \put(0.60151097,0.42371909){\color[rgb]{0,0,0}\makebox(0,0)[b]{\smash{LDV}}}%
    \put(0.04654943,0.00370678){\color[rgb]{0,0,0}\makebox(0,0)[rb]{\smash{$e_z$}}}%
    \put(0.09401236,0.03506203){\color[rgb]{0,0,0}\makebox(0,0)[lb]{\smash{$e_x$}}}%
    \put(0.04550973,0.08543379){\color[rgb]{0,0,0}\makebox(0,0)[rb]{\smash{$e_y$}}}%
  \end{picture}%
\endgroup

%% file: concl.tex
\section{Conclusions and outlook}
\label{sec:concl}

\begin{figure}[!ht]
  \centering
  \includegraphics[width=0.7\linewidth]{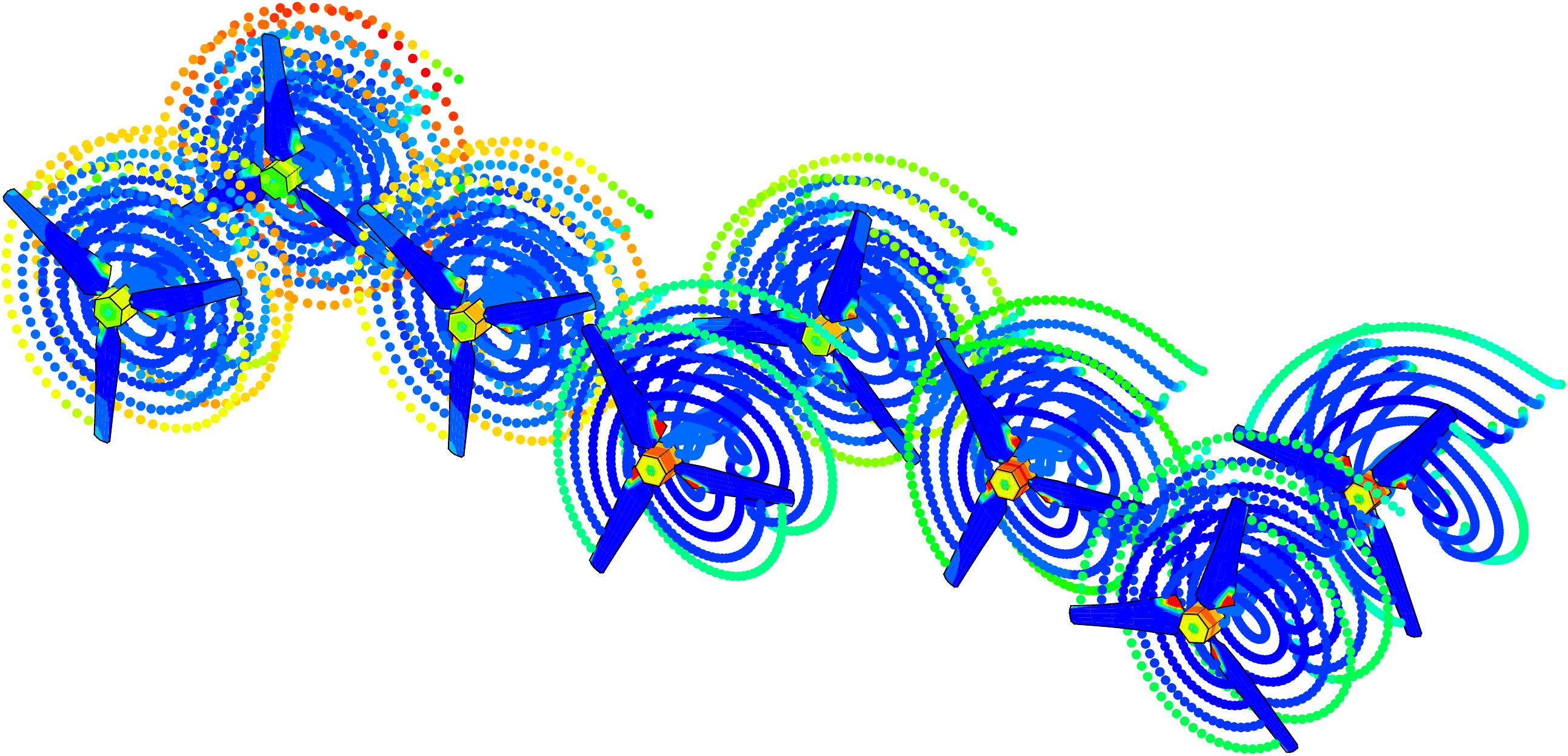}
    \caption{First coarse numerical simulation with eight turbines in close proximity.\label{fig:num_prospect}}
\end{figure}

A first study about interaction effects between two horizontal axis
marine current turbines has been successfully carried out.  On single
device configurations, the wake behind the turbine was characterised in
terms of velocity deficit and turbulence intensity, which showed that a
higher upstream ambient TI tends to reduce the wake influence
length. The behaviour of a single turbine has already been examined in
terms of performance, which lead  to determine that the range of TSR
yielding the most power was between three and four for this type of
blade geometry. These experimental data enabled the validation of the
numerical tool both on the wake computation and on the performance
evaluation. The comparison showed very promising results and we are
confident that the software will soon be able to model successfully multi-device
configurations (cf. figure~\ref{fig:num_prospect}). In addition, the
numerical software is currently being rewritten in order to be more
efficient and to include the latest numerical development such as a
better turbulence model.

Single configurations data have been used as a basis to carry out trials
on  two devices  configurations.  The study showed  that  wake interaction
effects between  the turbines exist  and that the downstream  turbine is
thus  deeply  affected  by  the  presence  of  an  upstream  device.   A
qualitative  and quantitative  characterization of  the  interaction has
been  presented concerning  both the  wake  and the  performance of  the
downstream turbine.  It is clear that increasing inter-device spacing to
retrieve higher  individual power can only  be done to  the detriment of
the total  number of  turbine rows  in a given  space.  So  a compromise
between individual  performance and the number of  energy converters has
to be made wisely when considering an array implantation.

Some of our prospects, which concern both the experimental and the
numerical aspect, consist in modelling other kinds of turbine prototypes
or in taking into account both wave and current effects on the behaviour
of marine current turbines. Recently, new performance trials were run
with a torque-meter instead of the load cell to measure the torque
directly on the hub. In addition, twin-turbine configurations were
tested in a flow with a 15\% TI to complement the study presented
here. Those results will be presented in~\cite{Mycek2013a, Mycek2013b}.


%% file: acknow.tex
\section*{Acknowledgment}

The authors would like to thank R\'egion Haute-Normandie for the
financial support granted for co-financed PhD theses and the CRIHAN
(Centre des Ressources Informatiques de HAute-Normandie) for their
available numerical computation resources.  We are also grateful to
Thomas Bacchetti, Jean-Matthieu Etancelin and Jean-Valery Facq for their
help in the present work. The authors would like to acknowledge the
scientific committee of EWTEC 2011 for selecting our paper to be
published in IJME.
